\documentclass[a4paper,11pt]{article}
\pdfoutput=1
\usepackage{jheppub,mathabx}

\usepackage{amsmath}                                      
\usepackage{amsthm}
\usepackage{physics}     
\usepackage{slashed}
\usepackage{cancel}
\usepackage{graphics}
\usepackage{multind}
\usepackage[latin1]{inputenc}
\usepackage{wrapfig}
\usepackage{amssymb}
\usepackage{comment}
\usepackage{latexsym}
\usepackage{mathtools}
\usepackage{dsfont}
\usepackage{mathrsfs}
\usepackage{hyperref}
\usepackage{pdfpages}
\usepackage{float}
\usepackage{graphicx}
\usepackage{framed}
\usepackage{amsfonts}
\usepackage{subfig}
\usepackage{enumitem}
\usepackage{color}
\usepackage{multirow,multicol}
\usepackage[table,xcdraw]{xcolor}
\usepackage[framemethod=TikZ]{mdframed}
\usepackage{xstring} 
\usepackage[normalem]{ulem}
\allowdisplaybreaks[4]

\newcommand{\s}{\,{\rm s}}

\newcommand{\GeV}{\,{\rm GeV}}
\newcommand{\TeV}{\,{\rm TeV}}
\newcommand{\cm}{\,{\rm cm}}

\def\circa#1{\,\raise.3ex\hbox{$#1$\kern-.75em\lower1ex\hbox{$\sim$}}\,}

\newcommand{\beq}{\begin{equation}}
\newcommand{\eeq}{\end{equation}}
\newcommand{\be}{\begin{equation}}
\newcommand{\ee}{\end{equation}}
\font\tenrsfs=rsfs10 at 12pt
\font\sevenrsfs=rsfs7
\font\fiversfs=rsfs5
\newfam\rsfsfam
\textfont\rsfsfam=\tenrsfs
\scriptfont\rsfsfam=\sevenrsfs
\scriptscriptfont\rsfsfam=\fiversfs
\renewcommand{\thesection}{\arabic{section}}

\newsavebox\MBox

\newcommand{\LDC}{\Lambda_{\textrm{DC}}}

\newcommand{\gdc}{g_{\textrm{DC}}}
\newcommand{\adc}{\alpha_{\scriptscriptstyle\textrm{DC}}}
\newcommand{\SU}{\,{\rm SU}}
\newcommand{\SO}{\,{\rm SO}}
\newcommand{\U}{\,{\rm U}}
\newcommand{\GDC}{ G_{\rm DC}}
\newcommand{\NDC}{N_{\rm DC}}
\newcommand{\GSM}{G_{\rm SM}}
\newcommand{\ie}{\emph{i.e. }}
\newcommand{\MQ}{M_\mathcal{Q}}
\newcommand{\MPL}{M_{\mathrm{Pl}}}
\newcommand{\Q}{\mathcal Q}

\newcommand{\DC}{{\scriptscriptstyle\text{DC}}}
\newcommand{\SM}{{\scriptscriptstyle\text{SM}}}

\numberwithin{equation}{section}
\usepackage{epsfig}
\allowdisplaybreaks[4]

\makeatletter
\def\hhref#1{\href{http://arxiv.org/abs/#1}{arXiv:#1}} 
\newcommand{\hhrefq}[1]{\IfSubStr{#1}{:}{\href{http://inspirehep.net/search?ln=en&ln=en&p=#1&of=hb&action_search=Search&sf=&so=d&rm=&rg=25&sc=0}{InSpires:#1}}{\hhref{#1}}}

\def\art{\@ifnextchar[{\eart}{\oart}}
\def\eart[#1]#2#3#4#5#6{{\rm #2}, {\em #3 \bf #4} {\rm (#6) #5} ({\em #1})}

\def\article{\@ifnextchar[{\earticle}{\oarticle}}
\def\oarticle#1#2#3#4#5#6{{\rm #1}, {\em ``#6''}, {\rm #2 #3 (#5) #4}}
\def\earticle[#1]#2#3#4#5#6#7{{\rm #2}, {\em ``#7''}, {\rm #3 #4 (#6) #5}  [\hhrefq{#1}]}
\def\hepart[#1]#2{{\rm #2, \em#1}}
\def\heparticle[#1]#2#3{#2, {\em ``#3''} [\hhrefq{#1}]}

\title{Gluequark Dark Matter}
\author{Roberto Contino$^{a,\,b}$, Andrea Mitridate$^{a,\,b}$, Alessandro Podo$^{a,\,b}$, Michele Redi$^c$}
   \affiliation{$^a$ Scuola Normale Superiore, Piazza dei Cavalieri 7, 56126, Pisa, Italy}
   \affiliation{$^b$  INFN sezione di Pisa, Italy}
     \affiliation{$^c$  INFN sezione di Firenze, Via G. Sansone 1; I-59100 Sesto F.no, Italy}
\abstract{We introduce the gluequark Dark Matter candidate, an accidentally stable bound state made of adjoint fermions and gluons from a new confining gauge force. Such scenario displays an unusual cosmological history where perturbative freeze-out is followed by a non-perturbative re-annihilation period with possible entropy injection. When the gluequark has electroweak quantum numbers, the critical density is obtained for masses as large as PeV. Independently of its mass, the size of the gluequark is determined by the confinement scale of the theory, leading at low energies to annihilation rates and elastic cross sections which are large for particle physics standards and potentially observable in indirect detection experiments.}

\mdfsetup{%
backgroundcolor=gray!10,   %
 skipabove = 10pt,               %
 skipbelow = 20pt,  		  %
 innertopmargin=20pt, 	  %
 innerbottommargin=20pt,   %
 linewidth = 0pt, 		  %
 innerleftmargin = 20pt,	  %
innerrightmargin = 20pt,       %
roundcorner=30pt            %
}

\newcounter{box}
\newcounter{ijk}
\setlist{  
  listparindent=\parindent,
  parsep=0pt,
}

\begin{document}
\maketitle

\section{Introduction}

The striking success of the Standard Model (SM) in reproducing laboratory tests of fundamental interactions and its failure to explain some of the
structural features of our Universe motivate the study of extensions based on its same principles of simplicity and elegance. 
From a modern point of view the SM is understood as an effective field theory with a very high ultraviolet cut-off, which appears renormalizable at
energies currently probed in experiments.
This feature notoriously gives rise to the SM hierarchy problem,  but is also at the very origin of the attractive properties of the SM.
In particular, global symmetries arise accidentally in the infrared and 
explain in the most economical way baryon and lepton number conservation, flavour and electroweak (EW) precision tests.

We consider these remarkable properties as paradigmatic, providing a compelling guidance to build possible extensions of the SM, even at the price of sacrificing 
the naturalness of the electroweak scale (as hinted anyway by experiments). 
In particular, the cosmological stability of Dark Matter (DM) can be elegantly explained in terms of  accidental symmetries, in analogy with the stability of the proton
following from baryon number conservation.
This has to be contrasted with SM extensions where global symmetries are imposed ad hoc, like for example the case of $R$-parity in supersymmetry.
A simple way to generate accidental symmetries is to extend the gauge theory structure of the SM by postulating a new confining dark color group.
In this paper we will continue the exploration initiated in Refs.~\cite{1503.08749,1707.05380}
of theories where the dark sector comprises new quarks transforming as real or vector-like representations under both the SM and dark gauge groups, 
and where the Higgs field is elementary. Such framework, also known as Vector-Like Confinement~\cite{0906.0577}, provides a safe non-trivial extension of the SM,
since it gives small and calculable deviations to EW observables that are in full agreement with current data and potentially observable 
at future colliders~\cite{Barducci:2018yer}. 

Previous studies of accidental DM focused on baryons or mesons of the dark dynamics as DM candidates,
see~\cite{1604.04627} for a review.
A systematic analysis of models with baryonic DM was performed in Refs.~\cite{1503.08749,1707.05380}.
In the regime where dark quark masses are below the dark color dynamical scale, $M_\Q < \LDC$, it was found that a correct relic abundance can be obtained for DM masses of order 100$\,$TeV. Lighter DM masses, down to $\mathcal{O}(10\,{\rm TeV})$, are instead allowed if $M_\Q > \LDC$. Mesonic DM candidates can be even lighter and can arise for example as pseudo Nambu-Goldstone bosons (NGBs) of a spontaneously broken global symmetry of the dark dynamics.

\begin{figure}[t]
\centering
\includegraphics[width=.35\textwidth]{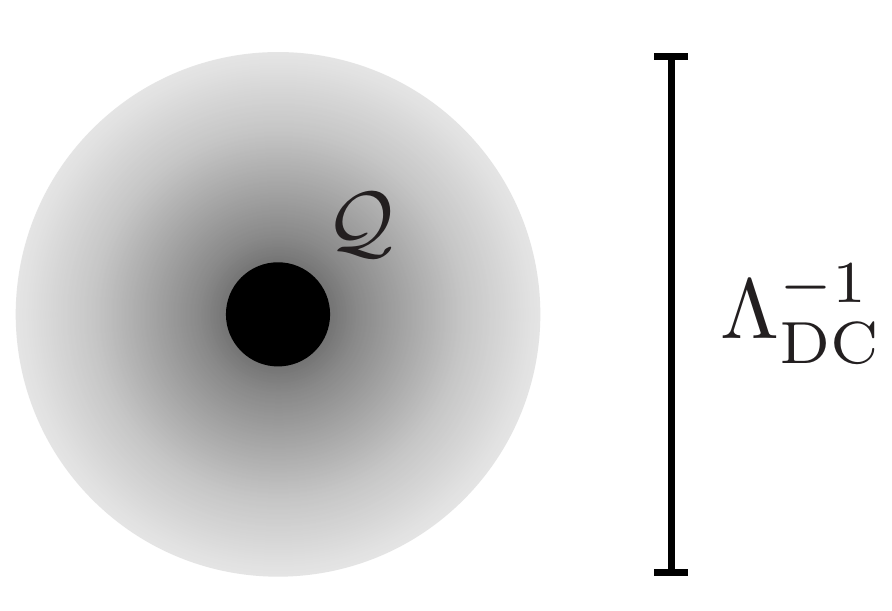}
\caption{\emph{Cartoon of the gluequark DM candidate. A heavy fermion in the adjoint of color gives rise to color singlet state 
surrounded by a gluon cloud of size $1/\LDC\gg 1/M_\Q$.}}
\label{fig:chicartoon}
\end{figure}

In this work we explore scenarios with a new kind of accidental composite DM candidate, the \emph{gluequark},
which has properties different from those of dark baryons and mesons in several respects. 
Gluequarks are bound states made of one dark quark and a cloud of dark gluons 
in theories where the new fermions transform in the adjoint representation of dark color. 
They are accidentally stable due to dark parity, an anomaly free subgroup of dark fermion number, which is exact
at the level of the renormalizable Lagrangian. 
Depending on the SM quantum numbers of the new fermions, violation of dark parity can arise from UV-suppressed dimension-6 operators thus ensuring
cosmologically stable gluequarks for sufficiently large cut-off scales.
Contrary to baryons and mesons, the physical size of the gluequark is determined by the confinement 
scale independently of its mass. In the regime of heavy quark masses, $M_\Q > \LDC$, this implies a physical size larger than its Compton wavelength,
see Fig.~\ref{fig:chicartoon}.
The annihilation cross section for such a large and heavy bound state can be geometric, much larger than the perturbative unitarity 
bound of elementary particles. This in turn modifies the thermal relic abundance and can lead to significant effects in indirect detection experiments.
Also, the resulting cosmological history is non-standard and different from that of theories with baryon or meson DM candidates.

Bound states made of one dark fermion and dark gluons were considered in Ref.~\cite{0908.1790}, where they couple to the SM sector
through the neutrino portal. Similar DM candidates were also studied in Refs.~\cite{Boddy:2014yra,Boddy:2014qxa}, in the context of 
supersymmetric gauge theories. There, bound states of one fermion (the dark gluino) and dark gluons arise as the partners of glueballs after 
confinement and were consequently called glueballinos.
Ref.~\cite{Boddy:2014yra} showed that the observed DM abundance can be reproduced by a mixture of glueballs and glueballinos provided that the dark and SM sectors are decoupled very early on in their thermal history. In such scenario the two sectors interact only gravitationally, the dark gluino being neutral under the SM gauge group.
Notice that the stability of glueballs in this case does not follow from an accidental symmetry but is a consequence of the feeble interaction
between the SM and dark sectors. In this paper we will focus on the possibility that dark fermions are charged under the SM gauge group, so that the lightest states of the dark sector may be accessible through non-gravitational probes. In this case the dark and visible sectors stay in thermal equilibrium until relatively low temperatures, of the order of $1$ GeV, and the thermal history of the Universe is rather different than that described in Refs.~\cite{Boddy:2014yra,Boddy:2014qxa}. In particular, we will argue that in our scenario dark glueballs cannot account for a sizeable fraction of DM because of BBN and CMB constraints.

Composite DM candidates from theories with adjoint fermions were also considered in the context of Technicolor models, see for example Refs.~\cite{Gudnason:2006ug,Kouvaris:2007iq}. Those constructions differ from ours in that technicolor quarks are assumed to transform as complex representations under the SM, but they can share common features with some of the models described in this paper~\footnote{Reference~\cite{Kouvaris:2007iq} for example considered gluequark DM in the context of the so-called Minimal Walking Technicolor model, but its estimate of the thermal relic abundance focuses on the perturbative freeze-out and does not include any of the non-perturbative effects described in this work.}.

The paper is organized as follows. Section \ref{sec:models} provides a classification of models with adjoint fermions that can lead to a realistic
DM candidate. We outline the cosmological history of the gluequark in section \ref{sec:cosmological}  and present our estimate for the thermal relic abundance in section \ref{sec:relicdensity}.  Section \ref{sec:bounds} discusses a variety of bounds stemming from cosmological and astrophysical data, DM searches at colliders, direct and indirect detection experiments. We summarize and give our outlook in section~\ref{sec:conclusions}. A discussion of the relevant cross sections can be found in the Appendices.

\section{The models}
\label{sec:models}

We consider the scenario in which the SM is extended by a new confining gauge group $\GDC$ (\emph{dark colour}),
and by a multiplet of Weyl fermions
$\Q$ (\emph{dark quarks}) transforming in the adjoint representation of $\GDC$ and as a (possibly reducible) 
representation $R$ under the SM group~$\GSM$:
\be
\Q \equiv \left( \mathrm{adj}, R \right) .
\ee
In particular, we consider models where the dark quarks have quantum numbers under $\SU(2)_L\times \U(1)_Y$ but are singlets of $\SU(3)_c$ color.
New fermions colored under $\SU(3)_c$ are an interesting possibility but are subject to very strong experimental constraints and their analysis 
deserves a separate study (see for example the recent discussion in Ref.~\cite{DeLuca:2018mzn}).
We assume $R$ to be a real or vector-like representation, so that the cancellation of $\GSM$ anomalies is automatic and mass terms for the dark quarks are allowed.

We performed a classification of the minimal models, \ie those with the smallest representations and minimal amount of fields, which give a consistent theory of DM. We refer to them as `minimal blocks'. Each block is characterised by two parameters: the dark quark mass $M_{\mathcal{Q}}$ and the value of the dark gauge coupling $\gdc$ at $M_{\mathcal{Q}}$. A $CP$-violating $\theta$ term can also be included but does not play an important role in what follows. 
The renormalizable lagrangian thus reads
\be
\label{eq:Lagrangian}
\mathcal{L} = \mathcal{L}_{\mathrm{SM}} - \dfrac{1}{4 \gdc^2} \mathcal{G}_{\mu\nu}^{2} + \mathcal{Q}^{\dagger} i \bar{\sigma}^{\mu} D_{\mu} \mathcal{Q} - \dfrac{M_{\mathcal{Q}}}{2} \left( \mathcal{Q}\mathcal{Q} + \mathcal{Q}^{\dagger}\mathcal{Q}^{\dagger}\right).
\ee
It is possible to combine more than one minimal block; in this case the number of parameters increases: each module can have a different mass and, depending on the SM quantum numbers, Yukawa couplings with the Higgs boson may be allowed. 

As long as all dark quarks are massive, the theory described by (\ref{eq:Lagrangian}) confines in the infrared forming bound states. The symmetry
$\mathcal{Q} \rightarrow -\mathcal{Q}$, \emph{dark parity},  is an
accidental invariance of the renormalizable
Lagrangian.
The physical spectrum is characterized by states that are either even or odd under dark parity. The gluequark, denoted by $\chi$ in the following, 
is the lightest odd state and has the same SM quantum numbers of its constituent dark quark, thus transforming as an electroweak multiplet. 
Radiative corrections will induce a mass splitting among different components, with the lightest state being accidentally stable at the renormalizable level thanks to its odd dark parity.  The mass difference computed in Ref.~\cite{Cirelli:2005uq} shows that the lightest component is always the electromagnetically neutral one, which therefore can be a DM candidate provided it has the correct relic density.

We select models with a suitable gluequark DM candidate by requiring them to be free of Landau poles below $10^{15}\,$GeV. This is a minimal assumption considering that, as discussed below, astrophysical and cosmological bounds on the gluequark lifetime can be generically satisfied only for a sufficiently large cut-off scale. It is also compatible with Grand Unification of SM gauge forces. The ultraviolet behaviour of each model is dictated by the number of dark colors $N_{\rm DC}$ and by the dimension of the SM representation $R$, i.e. by the number of Weyl flavors $N_f$. Models with too large $N_f$ or $N_{\rm DC}$ imply too low Landau poles for $\GSM$, and are thus excluded from our analysis. The list of minimal blocks that satisfy our requirements is reported in Table~\ref{tab:reps} for $\SU(N_{\rm DC})$, $\SO(N_{\rm DC})$ and ${\rm Sp}(N_{\rm DC})$ dark color groups.
\begin{table}[t]
\centering
\begin{tabular}{|c|c|c|c|c|c|c|c|}
\hline
\rowcolor[HTML]{C0C0C0} 
 &  Quantum numbers & \multicolumn{3}{|c|}{$\NDC$} &  Accidental &  & Classical \\ 
\rowcolor[HTML]{C0C0C0} 
\multirow{ -2}{*}{$N_f$} & ${\rm SU}(2)\times{\rm U}(1)$ & ${\rm SU}$ & ${\rm SO}$ & ${\rm Sp}$ & Symmetry & \multirow{ -2}{*}{$\mathcal{O}_{\rm dec}$} & $\left[\mathcal{O}_{\rm dec}\right]$ \\ \hline\hline
 1    &   $N\equiv1_0$   &All &All &All &   $\mathbb{Z}_2$   &   $\ell H \mathcal{G}_{\mu\nu}\sigma^{\mu\nu}N$     &   6   \\ \hline 
 3    &   $V\equiv3_0$   &   $\leq3$ &   $\leq4$ &   $1$   &   $\mathbb{Z}_2$   &   $\ell H \mathcal{G}_{\mu\nu}\sigma^{\mu\nu}V$   &   6   \\ \hline
 4    &   $L\equiv2_{1/2}\oplus\bar{L}\equiv 2_{-1/2}$   &   $\leq4$ &   $\leq6$ &   $\leq2$  &   ${\rm U}(1)$   &   $\ell \mathcal{G}_{\mu\nu}\sigma^{\mu\nu}L$   &   5   \\ \hline
 6    &   $T\equiv3_1\oplus\bar{T}\equiv 3_{-1}$   &   $ 2$ &   $3$ &   $1$  &   ${\rm U}(1)$   &   $\ell H^c \mathcal{G}_{\mu\nu}\sigma^{\mu\nu}T$   &   6  \\ \hline
\end{tabular}
\caption{\label{tab:reps} \it Minimal building blocks for models of gluequark DM. We require that a multiplet contains an electromagnetic neutral component and that the gauge couplings do not have Landau poles below $10^{15}\,\rm{GeV}$, assuming a representative mass of $100\,\rm{TeV}$ for the dark quarks. We denote with $\ell$ the SM lepton doublets.}
\end{table}
Each block is characterized by its accidental symmetry (that can be larger than the dark parity) and by the
dimensionality of the lowest-lying operator $\mathcal O_{\rm dec}$ which violates it. The latter has the form
\begin{equation}
\label{eq:Odec}
\mathcal{O}_{\rm dec} = \mathcal{O}_{\rm SM} \mathcal{G}^{a}_{\mu\nu}  \sigma^{\mu\nu} \mathcal{Q}^{a},
\end{equation}
where $\mathcal{O}_{\rm SM}$ is a SM composite operator matching the $\SU(2)_L\times \U(1)_Y$ quantum numbers of the dark quark $\mathcal{Q}$.
The operator (\ref{eq:Odec}) can in general induce the mixing of the gluequark with SM leptons, providing an example of partial compositeness.
As long as the theory is not in the vicinity of a strongly-coupled IR fixed point at energies $E \gg M_{\mathcal{Q}}, \LDC$, the dimension of $\mathcal{O}_{\rm dec}$
is simply given by $[\mathcal{O}_{\rm dec}] = 7/2 + [\mathcal{O}_{\rm SM}]$, as reported in the sixth column of Table~\ref{tab:reps}.
Among the minimal blocks, the $L\bar L$ model has $[\mathcal O_{\rm dec}]=5$ classically.
In this case the naive suppression of the gluequark decay rate is not enough to guarantee cosmological stability, 
although a stable DM candidate can still be obtained through additional dynamics, see the discussion in Appendix~\ref{app:LL}.
In the remaining minimal blocks the classical dimension of $\mathcal O_{\rm dec}$ is 6 and the gluequark can be sufficiently long lived.
Indirect detection experiments and data from CMB and 21 cm line observations set important constraints on these
models which will be discussed in section~\ref{sec:bounds}. 

The behaviour of the theory at energies above the confinement scale depends largely on the number of dark flavors $N_f$ and on the value of the dark coupling
$\gdc$ at the scale $\MQ$. One can identify two regimes. In the first, $\gdc(\MQ)$ is perturbative and this necessarily implies a confinement scale smaller than
the quark mass, $\LDC < \MQ$; we will call this the `heavy quark' regime. In this case, depending on the value of $N_f$, there are three possible behaviours.
For $N_f \geq N_f^{\rm AF} = 11/2$ the theory is not asymptotically free, hence starting from the UV the coupling gets weaker at lower scales until one reaches the quark mass 
threshold  below which the dynamics becomes strong and confines.~\footnote{Notice that the value of $N_f^{\rm AF}$, in the case of adjoint fermions, does not depend on the gauge group.}
For $N_f^c \leq N_f < N_f^{\rm AF}$, where $N_f^c$ is the non-perturbative edge of the conformal window, the theory flows towards an IR fixed point at low energies 
until the quark mass threshold is passed, below which one has confinement.
Finally, if $N_f < N_f^{c}$ the coupling grows strong quickly in the infrared and
confinement is triggered without delay. Only for this latter range of values of $N_f$ the confinement scale can be larger than the quark mass, $\MQ < \LDC$; we will refer to this as the `light quark' regime. The physical spectrum, the phenomenology and the thermal history are rather different in the two regimes.

The infrared behaviour of $\SU(\NDC)$ gauge theories with fermions in the adjoint representation was extensively studied through lattice simulations, 
see for example~\cite{Catterall:2011zf,Bergner:2016hip,Bergner:2017gzw,Athenodorou:2014eua,Bergner:2015adz,DeGrand:2015zxa,Ali:2016zke,Bergner:2017bky,Nogradi:2016qek,DeGrand:2013uha} and references therein.
There seems to be sufficient evidence for an infrared conformal phase of theories with $N_{\rm DC} =2$ colors and $N_f=4,3$ massless Weyl flavors, while results with $N_f =2$ are more uncertain though still compatible with a conformal regime.
Theories with $N_f=1$ are supersymmetric and have been shown to be in the confining phase.
The case with $N_{\rm DC}=3$ colors is much less studied and no firm conclusion can be drawn on the conformal window.
Notice that, independently of the number of colors, asymptotic freedom is lost for $N_f \geq 6$, while the existence of a weakly-coupled infrared fixed point
can be established for $N_f =5$ by means of perturbation theory.
Besides determining which phase the massless theory is in, simulations give also information on the spectrum of bound states.
In particular, information on the gluequark mass in the limit of heavy quark masses ($M_\Q \gg \LDC$) can be obtained from lattice simulations with
adjoint static sources, see for example Refs.~\cite{Foster:1998wu,Bali:2003jq}.

\paragraph{Heavy quark regime:}\label{subsec:heavy}
In the heavy quark regime, the lightest states in the spectrum are glueballs, while those made of quarks are parametrically heavier.
The value of the glueball mass is close to the one of pure gauge theories.
Lattice results for pure glue $\rm SU(3)$ theories show that the $0^{++}$ state is the lightest with mass $m_{0^{++}}\sim 7 \LDC$, see for example~\cite{Morningstar:1999rf}. 
Similar values are found for $\SU(\NDC)$ with different number of colors. The gluequark is expected to be the lightest state made of quarks, with a mass $M_\chi \sim \MQ$.
Other states made of more dark quarks (collectively denoted as mesons) quickly decay to final states comprising glueballs and gluequarks, 
depending on their dark parity.

The gluequark lifetime can be accurately estimated by computing the decay of its constituent heavy quark, similarly to
spectator calculations for heavy mesons in QCD. In the minimal blocks where the dark parity-violating operator has dimension~6
the main decay channel for the lightest gluequark $\chi^0$ is $\chi^0\to h \nu + n\Phi$ (where $\Phi$ indicates a glueball and $n\geq 1$). 
In the $V$ model of Table~\ref{tab:reps} with three dark flavors transforming as an EW triplet, the dim-6 operator
\begin{equation*}
\frac{g_{UV}^2}{\Lambda_{UV}^2} \left( H^{c \dagger} \sigma^i\ell \, \mathcal{G}_{\mu\nu}\sigma^{\mu\nu}\mathcal{Q}^i + h.c. \right)
\end{equation*}
induces the decay of the gluequark with inverse lifetime
\begin{equation}
\label{eq:lifetimeheavy}
\frac 1{\tau(\chi_0)}\simeq  \frac{g_{UV}^4}{4096 \pi^3} \frac {M_\Q^5}{\Lambda_{UV}^4}
\simeq 10^{-28} g_{UV}^4 \left(\frac {M_\Q}{100\, \rm TeV}\right)^5 \left(\frac{10^{18}\, \rm GeV}{\Lambda_{UV}}\right)^4 \, {\rm s}^{-1}\, .
\end{equation}
Similar results apply for the $N$ and $T\oplus\bar T$ minimal blocks.

Glueballs can decay to SM particles through loops of dark quarks. 
In particular, since the latter are assumed to have electroweak charges, glueballs can always decay to photons through dimension-8 operators
of the form $\mathcal{G}_{\mu\nu}\mathcal{G}^{\mu\nu} W_{\alpha\beta}W^{\alpha\beta}$ generated at the scale $\MQ$. For all the minimal models in Table \ref{tab:reps} this is the lowest-dimensional operator which induces glueball decay. 
The partial width into photons is determined to be~\cite{0903.0883,0911.5616}~\footnote{To derive this and the following decay rates 
we used the value of the matrix element $\langle0 \vert \mathcal{G}_{\mu\nu}\mathcal{G}^{\mu\nu} \vert\Phi \rangle$
computed on the lattice for $SU(3)$, see Ref.~\emph{e.g.}~\cite{0808.3151}.}
%
\begin{equation}
\label{eq:lifetime8}
\Gamma(\Phi\to \gamma \gamma)\simeq 0.7 \,{\rm s}^{-1} \left(\frac {N_{\rm DC}}{3}\right)^2\left( \frac{M_{\Phi}}{500\GeV}\right)^9 
\left( \frac{100\TeV}{\MQ}\right)^8.
\end{equation}
When phase space allows, the decay channels $Z\gamma$, $W^+W^-$ and $ZZ$ open up producing one order-of-magnitude smaller lifetime.
Relatively long-lived glueballs, as implied by the estimate~(\ref{eq:lifetime8}), are subject to cosmological and astrophysical constraints as discussed
in section~\ref{sec:bounds}.

Models with Yukawa couplings to the Higgs doublet can be obtained combining minimal blocks. 
In this case 1-loop radiative effects at the scale $\MQ$ generate the dimension-6 effective operator $|H|^2 \mathcal{G}_{\mu\nu}^2$, 
inducing a much shorter lifetime. 
If their mass is high enough, $M_{\Phi}>2 m_h$, glueballs predominantly decay to two Higgs bosons with a decay width
\begin{equation}
\Gamma(\Phi\to hh)\simeq 10^{12} \,{\rm s}^{-1} \left(\frac{N_{\rm DC}}{3}\right)^2\left(\frac{y_1 y_2}{0.1}\right)^2 
\left(\frac{M_{\Phi}}{500\GeV}\right)^5 \left( \frac{100\TeV}{\bar M_{\Q} }\right)^4,
\end{equation}
where $\bar M_\Q = (M_{\Q_1} M_{\Q_2})^{1/2}$, and $y_{1,2}$, $M_{\Q_{1,2}}$ are respectively the Yukawa couplings and masses of the dark quarks circulating in the loop. Lighter glueballs can decay through the mixing with the Higgs boson; as for the Higgs, the dominant channel for $M_{\Phi}< 150\,$GeV is that into bottom quarks,  with a corresponding partial width~\footnote{The scaling $\Gamma(\Phi\to b\bar b) \sim M_\Phi^{7}$ is approximately correct for $M_\Phi \ll m_h$, though eq.~(\ref{eq:taubb}) is a good numerical estimate for $m_h \sim M_\Phi < 150\GeV$ as well.}
\begin{equation}
\label{eq:taubb}
\Gamma(\Phi\to b\bar b)\simeq 3 \cdot10^{7}\,{\rm s}^{-1} \left(\frac{N_{\rm DC}}{3}\right)^2\left(\frac{y_1 y_2}{0.1}\right)^2 
\left( \frac{M_{\Phi}}{50\GeV}\right)^7 \left( \frac{10\TeV}{\bar M_\Q}\right)^4.
\end{equation}

\paragraph{Light quark regime:}\label{subsec:light}
If dark quarks are lighter than $\LDC$, the physical spectrum is radically different and one expects spontaneous breaking of the global $SU(N_f)$ symmetry down to $\SO(N_f)$. The lightest states are thus the (pseudo) Nambu-Goldstone bosons $\varphi$, while the DM candidate is the gluequark, accidentally stable and with a mass of the order of the confinement scale~$\LDC$. As discussed in section~\ref{sec:relicdensity}, and similarly to the baryonic DM theories of Ref.~\cite{1503.08749}, reproducing the correct DM relic density in this regime fixes $\LDC \sim 50\TeV$. The NGBs with SM quantum numbers get a mass from 1-loop electroweak
corrections, which is predicted to be ${\mathcal O}(10\,\text{TeV})$ for the value of $\LDC$ of interest.
Besides such a radiative correction, the quark mass term breaks explicitly the $\SU(N_f)$ global symmetry and gives an additional contribution. Including both effects, the NGB mass squared is given by
\begin{equation}
\label{eq:pionmass}
m_\varphi^2 = c_0 M_\Q \LDC + c_1 \frac {3\alpha_2}{4\pi} I(I+1)\LDC^2\, ,
\end{equation}
where $I$ is the weak isospin of the NGB and $c_{0,1}$ are ${\mathcal O}(1)$ coefficients.

For fermions in the adjoint representations, only models with $N_f<5$ light quarks can be in the regime $\MQ < \LDC$, since those with more fermions are either IR conformal or IR free. Therefore, among the minimal blocks of Table~\ref{tab:reps} only two are compatible with the light quark regime, \ie the $V$ model and the $L\oplus\bar{L}$ model. 
The $V$ model has a global symmetry breaking $\SU(3) \rightarrow \SO(3)$ which leads to five NGBs transforming 
as an electroweak quintuplet. In the $L\oplus \bar{L}$ model one has $\SU(4) \rightarrow \SO(4)$ and nine NGBs transforming as 
$3_\pm, 3_0$ of $\SU(2)_{\rm EW}\times \U(1)_Y$. 

The limit of very small quark masses, $\MQ\LDC \ll m_\varphi^2$, is experimentally interesting, since NGBs have predictable masses.
In general, the lightest NGBs decay to SM final states through  anomalies or Yukawa couplings, as in the case of the $V$ model.
In some cases, however, some of the NGBs are accidentally stable due to unbroken symmetries of the renormalizable Lagrangian. An explicit violation of such accidental symmetries is expected to arise from higher-dimensional operators, possibly resulting into long-lived particles.
An example of this kind is given by the $L\oplus\bar L$ model, where NGBs made of $LL$ or $\bar L\bar L$ constituents have $U(1)$ number $\pm 2$ and are stable at the renormalizable level, see Appendix~\ref{app:LL}.

Since we assumed the dark quarks to transform as real or vectorlike representations under the SM gauge group, the fermion condensate responsible for the
global symmetry breaking in the dark sector can be aligned along an ($SU(2)_L \times U(1)_Y$)-preserving direction (in non-minimal models, Yukawa interactions can generate a vacuum misalignment leading to Higgs partial compositeness, see~\cite{Kaplan:1983fs,1508.01112}).
As the strong dynamics preserves the EW gauge symmetry of the SM, it also affects electroweak precision observables through suppressed corrections which are easily compatible with current constraints for sufficiently high
values of $\LDC$, as required to reproduce the DM relic abundance~\cite{0906.0577,1707.05380,Barducci:2018yer}.

Besides the NGBs, the physical spectrum comprises additional bound states with mass of order $\LDC$.
These include the gluequarks, which are expected to be the lightest states with odd dark parity, and mesons (i.e. bound states made of more than 
one dark quark)~\footnote{The existence of stable baryons in theories with adjoint fermions was investigated in Refs.~\cite{Bolognesi:2007ut,Auzzi:2008hu}, 
where stable skyrmion solutions were identified and conjectured to correspond to composite states with mass of $O(N_{\rm DC}^2)$, interpolated from the vacuum by 
non-local operators.  We will not include these hypothetical states in our analysis. In the light quark regime they are expected to annihilate with a geometric cross section 
and contribute a fraction of DM relic
density comparable to that of the gluequarks.}.
Except for the lightest gluequark, which is cosmologically stable, all the other states promptly decay to final states comprising NGBs and gluequarks, depending on their dark parity. 
In the minimal blocks where dark parity is broken by the dimension-6 operator $\ell H \mathcal{G}_{\mu\nu}\sigma^{\mu\nu}\mathcal{Q}$, the
most important decay channels of the gluequark are $\chi^0 \to h\nu$ and $\chi^0\to h\nu + \varphi$. 
The two-body decay dominates at large-$N_{\rm DC}$ and gives a lifetime of order
\begin{equation}
\label{eq:lifetimelight}
\frac 1{\tau(\chi_0)}\sim  \frac{g_{UV}^4}{8\pi} \frac {M_\chi^3 f_\chi^2}{\Lambda_{UV}^4}
= 4\! \times\! 10^{-26} g_{UV}^4 \left[\frac {M_\chi}{100\, \rm TeV}\right]^3 \left[\frac {f_\chi}{25\, \rm TeV}\right]^2
 \left[\frac{10^{18}\, \rm GeV}{\Lambda_{UV}}\right]^4  {\rm s}^{-1}
\end{equation}
where $f_\chi$ is the decay constant of the gluequark~\footnote{This has been defined by 
$\langle 0 | {\cal G_{\mu\nu}} \sigma^{\mu\nu} \Q | \chi (p,r) \rangle = f_\chi M_\chi u_r(p)$, and scales as 
$f_\chi \sim M_\chi (N_{\rm DC}/4\pi)$ in the large-$N_{\rm DC}$ limit.}.

\vspace{0.5cm}
To summarize our discussion on models, Table~\ref{tab:reps} reports the minimal blocks which have a potentially viable DM candidate and a sufficiently
high cut-off, above $10^{15}\GeV$, as required for SM Grand Unification and to suppress the DM decay rate.
In particular, the requirement on the absence of Landau poles restricts the list of possible models to a few candidates.
As mentioned before, the case of the singlet was studied already in the literature~\cite{Boddy:2014yra, Boddy:2014qxa}, and it will not be considered further in this work. We find that in all the other minimal blocks of Table~\ref{tab:reps} the $SU(3)_c\times SU(2)_L\times U(1)_Y$ gauge couplings unify with much lower 
precision than in the SM. Making the dark sector quantitatively compatible with SM Grand Unification thus requires extending these minimal blocks by including additional matter fields. Also, it would be interesting to explore the possibility of unifying both the visible and dark gauge couplings.
We leave this study to a future work.

In the next sections we will discuss the thermal history of the Universe and try to estimate the DM relic density: section 3 explains the general mechanisms at work and is largely independent of the details of the models; section 4 gives a concrete example, adopting as a benchmark the $V$ model of Table~\ref{tab:reps}, i.e. the minimal block with an $SU(2)_L$ triplet. For a discussion of the $L \oplus\bar{L}$ model see Appendix~\ref{app:LL}.


\section{Cosmological History} 
\label{sec:cosmological}

The Universe undergoes different thermal histories in the light and heavy quark regimes.
We first give a brief overview of such evolution, followed by a more detailed discussion with quantitative estimates.

In the light quark regime the thermal history is relatively simple and similar to that described for baryonic DM in Ref.~\cite{1503.08749}. 
Dark color confines when dark quarks are relativistic and in thermal equilibrium. After confinement
 the gluequarks annihilate into NGBs with a non-perturbative cross section $\sigma v_{\rm rel}\sim \pi/\LDC^2$, while glueballs are heavy and unstable. 
At  temperatures $T\sim M_\chi/25$ the annihilation processes freeze out and the gluequarks start behaving as ordinary thermal relics.

In the heavy quark regime the thermal history is more complex and characterized by three different stages.
Before confinement ($T\gtrsim\LDC$), free dark quarks annihilate into dark gluons and undergo perturbative freeze-out at $T\sim \MQ/25$ 
(see section~\ref{subsec: pertfo}).
At confinement ($T\sim\LDC$), the vast majority of the remaining dark quarks hadronizes into gluequarks, while the plasma of dark gluons is 
converted into a thermal bath of non-relativistic glueballs. The formation of mesons is suppressed by the low density of dark quarks compared to the ambient dark gluons. Glueballs overclose the Universe if they are cosmologically stable, therefore we consider the region of the parameter space where their lifetime is sufficiently short.
As first pointed out in~\cite{Scherrer:1984fd,Kamionkowski:1990ni,McDonald:1989jd}, and recently reconsidered in~\cite{hep-ph/0005123,1506.07532,1811.03608}, decays of non-relativistic particles with a large and non-thermal energy density -- like  the glueballs  -- 
can modify the standard relation between the scale factor and the temperature during the cosmological evolution. 
If the glueballs are sufficiently long lived and dominate the energy density of the Universe at some stage of the cosmological evolution, 
the standard scaling $a\varpropto T^{-1}$ is modified into $a\varpropto T^{-8/3}$. During this early epoch of matter domination, the Universe expands faster than in the radiation-dominated era, leading to an enhanced dilution of the DM relic density (see section \ref{subsec: dil}). 
Finally, interactions among gluequarks can lead to a second stage of DM annihilation through the process
\be\label{eq:recomb}
\setlength{\jot}{1.5pt}
\begin{split}
\chi+\chi\rightarrow \Q&\Q^*+\Phi/V\\ 
& \raisebox{2pt}{$\drsh$}\,\Q\Q\rightarrow\,{\rm SM}
\end{split}
\ee
where $\Q\Q^*$ is an excited bound state of dark quarks and $V$ stands for a SM vector boson or possibly a Higgs boson in models with Yukawa interactions. 
An analogous mechanism was first discussed in Ref.~\cite{hep-th/0405159,hep-ph/0611322,0712.2681} and more recently by Ref.~\cite{1606.00159,DeLuca:2018mzn,1802.07720}.
The process \eqref{eq:recomb} proceeds in two steps. Initially, an excited bound state $\Q\Q^*$ with size $\mathcal{O}( 1/\LDC)$ 
is formed by a collision of two $\chi$'s through a recombination of the constituent heavy quarks. This is similar to what happens 
for example in hydrogen anti-hydrogen scattering~\cite{Morgan:1970yz}.
As a consequence of the large size of the gluequark (see the discussion in section~\ref{subsec:reann}), 
the corresponding recombination cross section is expected to be large $\sigma_\text{rec} \approx \pi/\LDC^2$.
Once formed, the $\Q\Q^*$ can either decay ($\Q\Q^*\rightarrow \Q\Q + V\rightarrow \rm{SM} $) or be dissociated back into two gluequarks 
by interactions with the SM and glueball baths ($\Phi/ \,V+\Q\Q\rightarrow \chi+\chi$). A naive estimate shows that 
the latter process typically dominates. This is because the largest contribution to the total cross section comes from scatterings with 
large  impact parameters, $b\sim1/\LDC$, in which the $\Q\Q^*$ is produced with a large angular momentum,  $\ell\sim M_\Q v b$. 
Bound states with $\ell \gg 1$ take more time to de-excite to lower states, and dissociation can happen before they reach the 
ground state. The annihilation of gluequarks through recombination is therefore inefficient as long as the glueball bath is present. 
Only when the glueballs decay away, a second stage of DM annihilation can take place through the process \eqref{eq:recomb}.


\subsection{Thermal freeze-out}\label{subsec: pertfo}

Thermal freeze-out is the first (only) phase of the cosmological evolution in the regime with heavy (light) quarks. 
In this stage the number density of free dark quarks (for $\MQ > \LDC$) or of gluequarks (for $\MQ < \LDC$) 
is reduced until it becomes so low that chemical equilibrium is no longer attained and freeze-out takes place.
The number density at freeze-out is approximately given by
\begin{equation}
\label{eq:freezeout}
n(T_{\rm f.o.}) \simeq \frac{H(T_{\rm f.o.})}{\langle \sigma_{\rm ann} v_{\rm rel}\rangle}\, ,
\end{equation}
where $H$ is the Hubble parameter, and afterwards it is diluted by the Universe expansion.

In the heavy quark regime, free dark quarks annihilate with a perturbative cross section into dark gluons and into pairs of SM particles 
(vector bosons, Higgs bosons and fermions).
The freeze-out temperature is of order $T_{\rm f.o.}\approx \MQ/25$.
A general expression for the annihilation cross section is reported in Appendix~\ref{app:ann}, see eq.~(\ref{eq:annDCSM}).
For the $V$ model with $N_{DC}=3$ analysed in the next section, the annihilation cross section into dark gluons and SM fields is
\be\label{eq:pertann}
\langle\sigma_{\rm ann} v_{\rm rel}\rangle =\frac{\pi\adc^2}{M_\Q^2}\left(\frac{27}{96}+\frac 1 8 \left(1+\frac {25} {12}\right)\frac{\alpha_2^2}{\adc^2}+\frac 1 2 \frac{\alpha_2} {\adc}\right)\left(\frac{1}{6}S_3+\frac{1}{3}S_{3/2}+\frac{1}{2}S_{-1}\right)\,,
\ee
neglecting the mass of final states.
The term from annihilation into SM particles separately shows the contribution of vectors and fermions plus longitudinal gauge bosons.

Terms in the second parenthesis encode the Sommerfeld enhancement from dark gluon exchange:
$S_3$, $S_{3/2}$, $S_{-1}$ refer respectively to the 1, 8 and 27 color channels and are given by \cite{1402.6287,1702.01141}
\be
S_n=\frac {\adc} {v_{\rm rel}}\frac{2\pi n}{1-e^{-2\pi n\adc/v_{\rm rel}}}\,.
\ee

In the light quark regime gluequarks annihilate into NGBs with a cross section that is expected to scale naively as \be\label{eq:sigmaNP}
\langle \sigma_{\rm ann} v_{\rm rel}\rangle\sim \frac{\pi}{\LDC^2}\, ,
\ee
in analogy with nucleon-nucleon scattering in QCD \cite{Zenoni:1999st}. Nambu-Goldstone bosons are unstable and later decay into SM particles.


\subsection{Dilution}\label{subsec: dil}
\begin{figure}[t]
\centering
\includegraphics[width=.46\textwidth]{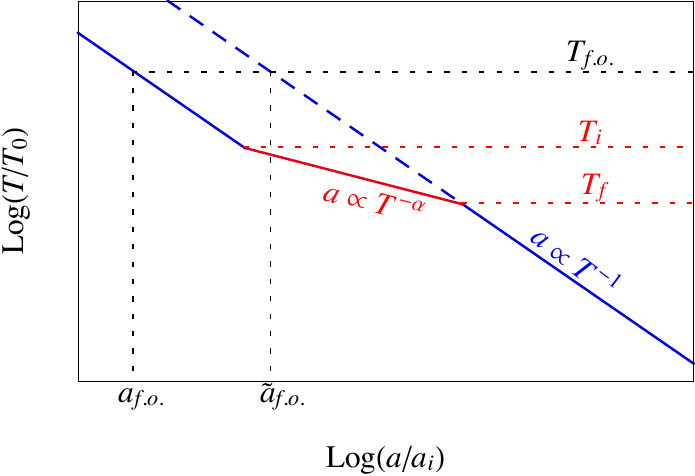}\quad
\includegraphics[width=.49\textwidth]{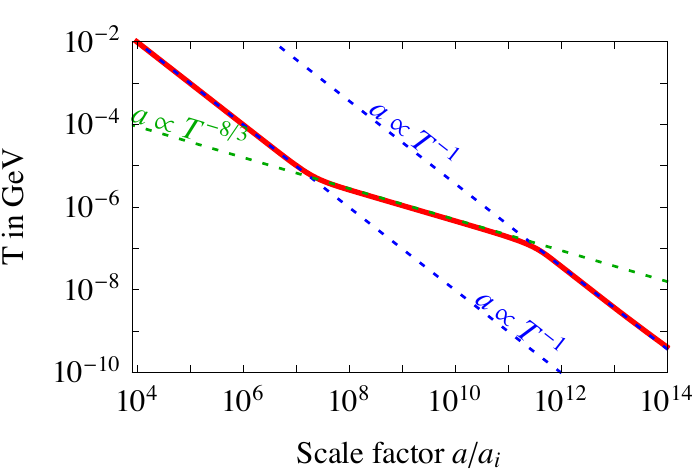}
\caption{\it {\bf Left:} Sketch of the non-standard $a(T)$ scaling. Values of $\alpha>1$ in eq.(\ref{eq:appdil}) imply a scale factor at freeze-out, $a_{f.o.}$, 
smaller than the one obtained from a standard cosmology, $\tilde{a}_{f.o.}$. This in turn leads to a suppression of the relic density by a factor
$(a_{f.o.}/\tilde{a}_{f.o.})^3=(T_f/T_i)^{3\alpha-3}$. {\bf Right:} Scaling obtained by solving numerically \eqref{eq:densityevol} and \eqref{eq:Friedmann} 
for $\MQ=100\TeV$ and $ T_{dc }=\LDC=10\GeV$. There exists an epoch during which the $a(T)$ scaling is very well approximated by a power law 
$a\varpropto T^{-\alpha}$  with $\alpha=8/3$.}
\label{fig:scaling}
\end{figure}

As well known, the number density of DM particles today is related to the number density at freeze-out by
\be
n_{\rm DM}(T_0)=n_{\rm DM}(T_{\,\rm f.o})\left(\frac{a_{\,\rm f.o.}}{a_0}\right)^3\,.
\ee
This relation is usually rewritten in terms of temperatures assuming that between freeze-out and today the standard scaling $a\varpropto T^{-1}$ holds. 
However, the validity of the standard scaling relies upon the assumption that entropy is conserved in the SM sector, i.e. that no energy
is injected into the SM plasma.
In presence of large entropy injection one can have an epoch during which $a$ grows faster than $a\varpropto T^{-1}$. 
In this case the relation between $n_{\rm DM}(T_0)$ and $n_{\rm DM}(T_{\rm f.o})$ is given by:
\be\label{eq:dil}
n_{\rm DM}(T_0)=n_{\rm DM}(T_{\,\rm f.o})\left(\frac{T_0}{T_{\,\rm f.o.}}\right)^3
\left(\frac{a\left(T_i\right)}{a\left(T_f\right)}\,\frac{T_i}{T_f}\right)^3\, ,
\ee
where $T_i$ and $T_f$ defines the temperature interval during which the non-standard scaling holds (see Fig.~\ref{fig:scaling}). 
The last term in parenthesis accounts for the suppression with respect to the naive relict density which would
be obtained using the standard scaling.
In the following we will show that late-time decays of dark glueballs can give rise to a non-standard scaling of the form $a\varpropto T^{-\alpha}$  with $\alpha>1$. The corresponding suppression factor thus reads:
\be\label{eq:appdil}
{\cal F} \equiv \left(\frac{a\left(T_i\right)}{a\left(T_f\right)}\,\frac{T_i}{T_f}\right)^3 = \left(\frac{T_f}{T_i}\right)^{3\alpha-3}\, .
\ee

After dark color confinement, the energy density of the Universe can be divided into a relativistic component, $\rho_R$, containing all the SM relativistic 
particles, and a non-relativistic one, $\rho_M$, containing all the dark-sector long-lived degrees of freedom (i.e. dark glueballs and gluequarks).
In particular, the energy density of glueballs at confinement is much larger than the corresponding thermal energy density for a non-relativistic species, 
and this can lead to an early epoch of matter domination. Neglecting the subleading contribution of gluequarks to $\rho_M$, the evolution of $\rho_{M, R}$ is governed by~\footnote{Here we omit the contribution of glueball annihilations into SM vector bosons to the evolution of $\rho_{M,R}$. This contribution is negligible in the region of the parameter space where dilution is sizeable. Both the rate of glueball annihilation at temperatures of order $\LDC$ and the glueball decay rate scale as $\LDC^9/\MQ^8$, so that the former, similarly to the latter, is expected to be smaller than Hubble when dilution is relevant. 
We have checked this naive expectation by verifying that after confinement the estimated annihilation rate is smaller than Hubble
on branch 1 of Fig.~\ref{fig:relic}.
}
\be\label{eq:densityevol}
\left\{
\begin{split}
\dot{\rho}_M =&-3H\rho_M-\Gamma_\Phi\,\rho_M\\
\dot{\rho}_R =&-4H\rho_R+\Gamma_\Phi\,\rho_M
\end{split}
\right.
\ee
where $\Gamma_\Phi$ is the glueball decay rate and the Hubble parameter $H$ is given by the Friedmann equation:
\be\label{eq:Friedmann}
H^2=\frac{8\pi G}{3}\left(\rho_R+\rho_M\right)\,.
\ee
Since in the relevant region of the parameter space the dark and SM sectors are in thermal equilibrium at dark confinement, 
the initial conditions at $T = T_{dc} \approx \LDC$ are given by 
\be\label{eq:rhoTrel}
\rho_{M}(T_{dc})=\xi\, \rho_{R}(T_{dc}) \quad\quad\textrm{ with }\quad\quad \xi\equiv\frac{g_D(T_{dc})}{g_*(T_{dc})}\,,
\ee
where $g_D(T)$ and $g_*(T)$ count the number of relativistic degrees of freedom in the dark and SM sector respectively. Furthermore, assuming that the 
decay products thermalize fast enough, the temperature of the Universe below $T_{dc}$ is related to the relativistic energy density by:
\be
\rho_R\equiv\frac{\pi^2 }{30}g_*\,T^4 \qquad (T < T_{dc})\,.
\ee

The evolution during the early matter-dominated epoch, if the latter exists, can be described by solving analytically eq.~(\ref{eq:densityevol}) 
at leading order in $\rho_R/\rho_M$ for cosmic times $t \ll 1/\Gamma_\Phi $~\cite{1506.07532}:
\begin{subequations}
\begin{align}
\rho_M&=\bar\rho_{M}\left(\frac{\bar a}{a}\right)^3e^{-\Gamma_\Phi(t-\bar t)}\\ \label{eq:radevol}
\rho_R& \simeq \bar\rho_{R}\left(\frac{\bar a}{a}\right)^4+\frac{2}{5}\,\sqrt{\frac{3}{8\pi}}\Gamma_\Phi\,M_{\rm Pl}\,\bar\rho_{M}^{1/2}\left[\left(\frac{\bar a}{a}\right)^{3/2}-\left(\frac{\bar a}{a}\right)^4\right]\,.
\end{align}
\end{subequations}
Here $\bar\rho_{M,R}$ and $\bar a$ denote the initial conditions at some time $\bar t$ much after the beginning of the matter-dominated epoch.
The relativistic energy density is given by the sum of $\bar\rho_{R}$ (first term in eq.\eqref{eq:radevol}), diluted as $a^{-4}$, and the energy injected by glueball decays (second term in eq.\eqref{eq:radevol}), diluted as $\sim a^{-3/2}$. Initially the first term dominates and the standard scaling $a \varpropto T^{-1}$ is 
obtained; as long as the glueball lifetime is long enough, the second term will start to dominate at some temperature $T_i$, implying a non-standard
scaling $a\varpropto T^{-8/3}$ (see Fig. \ref{fig:scaling}).
The value of $T_i$ can be found by equating the first and second terms of eq.\eqref{eq:radevol} and by 
using eqs.\eqref{eq:Friedmann},(\ref{eq:rhoTrel}):
\be\label{eq:Ti}
T_i\simeq T_{dc}\,\xi\times \left[\frac{\Gamma_\Phi \MPL}{4.15 \sqrt{g_*}\,T_{dc}^2\,\xi^2+\Gamma_\Phi \MPL}\right]^{2/5}\,.
\ee
The non-standard scaling ends when almost all the glueballs are decayed, \ie around $(t-t_{dc})\sim \Gamma_\Phi^{-1}$, where $t_{dc}$ is the time
at dark confinement. Using eqs. \eqref{eq:Friedmann} and \eqref{eq:rhoTrel}, one can translate this condition in terms of a temperature finding:
\be\label{eq:Tf}
T_f\simeq\sqrt{\MPL \Gamma_\Phi}\,.
\ee
From eq.\eqref{eq:appdil} it follows that late-time decays of glueballs dilute the naive relic density by a factor 
\be\label{eq:dilution}
\mathcal{F} = \left(\frac{T_f}{T_i}\right)^{5} =\frac{0.28}{g_*^{5/4}}\,\frac{\MPL^{5/2}\Gamma_\Phi^{5/2}}{T_{dc}^5\,\xi^5}\left(\frac{4.15 \sqrt{g_*}\, T_{dc}^2\,\xi^2+\Gamma_\Phi\MPL}{\Gamma_\Phi\MPL}\right)^2\,,
\ee
where ${\cal O}(1)$ numerical factors omitted in eq.\eqref{eq:Tf} have been included. When the glueballs are sufficiently long lived to give a sizeable dilution, the second term in the numerator inside the parenthesis of eq.\eqref{eq:dilution} can be neglected and $\mathcal{F}$ is very well approximated by:
\be\label{eq:dilution2}
\mathcal{F}\simeq \frac{4.82}{g_*^{1/4}}\, \frac{\sqrt{\MPL\Gamma_\Phi}}{T_{dc}\,\xi}\,.
\ee
While the analytic formulas (\ref{eq:radevol})-(\ref{eq:dilution2}) turn out to be quite accurate, in our estimate of the  relic density performed in section~\ref{sec:relicdensity} we will solve eq.~(\ref{eq:densityevol}) numerically without making any approximation.


\subsection{Reannihilation}\label{subsec:reann}

At $T = T_{dc}\sim \LDC$ the theory confines and the dark degrees of freedom reorganize into singlets of dark color. 
In the heavy quark regime,  the number density of gluons is much larger than the one of fermions and the vast majority of free quarks
$\Q$ hadronize into gluequarks.
These can then collide and recombine in excited $\Q\Q^*$ states by emitting an electroweak gauge boson 
(or a Higgs boson in theories with Yukawa couplings) or a glueball when kinematically allowed, see eq.~(\ref{eq:recomb}).
The process goes through a recombination of the constituent heavy quarks, while the direct annihilation of these latter has
a small and perturbative rate.
Given that gluequarks have a size of order $1/\LDC$, one expects naively a recombination cross section
of order $\sigma_{\rm rec}\sim 1/\LDC^2$. This value can in fact be reduced by kinematic constraints and the actual 
total cross section depends ultimately on the temperature at which the process takes place.
A detailed discussion and estimates for the recombination cross section are given in Appendix~\ref{app:supp}.

\begin{figure}[t]
\centering
\includegraphics[width=0.95\textwidth]{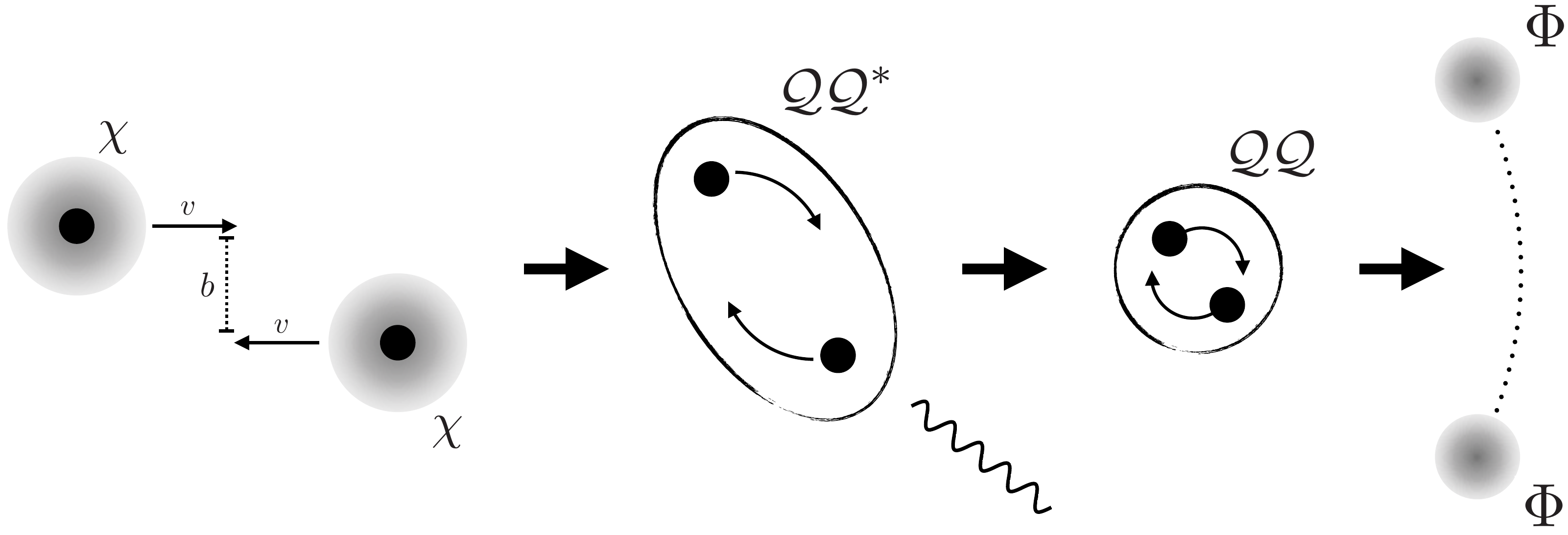}
\caption{\emph{Cartoon of the re-annihilation processes occurring after dark confinement. 
First, free quarks combine into color singlets gluequarks. Next, fast collisions
form excited $\Q\Q^*$ states that at sufficiently low temperatures fall into the ground state and decay.}}
\label{fig:reannihilation_cartoon}
\end{figure}

Once formed, $\Q\Q^*$ states with mass $M(QQ^*) > 2 M_\Q$ will promptly decay back to two gluequarks. Lighter states,
on the other hand, 
can either de-excite and thus decay into SM particles through the emission of a SM vector boson 
or a glueball ($\Q\Q^*\rightarrow \Q\Q + V/\Phi \rightarrow   \rm{SM}$), or be dissociated by interactions with the glueball and SM 
plasmas  ($\Phi/\, V+\Q\Q^*\rightarrow \chi+\chi$), see Fig.~\ref{fig:reannihilation_cartoon}.

If de-excitation occurs faster than dissociation, a second era of efficient DM annihilation can take place, 
reducing the gluequark number density. 
While re-annihilation processes  can be active over a long cosmological time interval, it is the last stage during which
the re-annihilation cross section gets its largest value $\sigma_{\rm rea}$ that is most important to determine the final gluequark
density. This last stage happens relatively quickly and can be characterized by a re-annihilation temperature $T_R$.
The exact value of $T_R$ depends on the rate of dissociation and is difficult to estimate.
The largest uncertainties arise from the calculation of the de-excitation rate, which can vary over several orders of magnitude.
We performed a thorough analysis taking into account the many dynamical ingredients which play a role in determining 
both the re-annihilation cross section and temperature. A detailed account is reported in Appendix~\ref{app:supp}.
We find that, under the most reasonable assumptions, dissociation of the most excited $QQ^*$ states occurs faster than de-excitation,
as long as the glueball bath originating from dark gluons confinement is present; therefore, the re-annihilation temperature is approximately equal to the one at which glueballs decay ($T_R \approx T_D$). Besides this most probable scenario, in the following we will also consider the other extreme possibility
where re-annihilation occurs right  after confinement ($T_R=\LDC$). The comparison between these two opposite scenarios will
account for the theoretical uncertainties intrinsic to the determination of the non-perturbative dynamics characterizing our 
DM candidate.

In both benchmark scenarios considered above the last stage of the re-annihilation epoch occurs while entropy is conserved in the Universe and can thus be described by a set of standard Boltzmann equations given in eq.~\eqref{eq:lateann}. They reduce to a single equation for sufficiently large de-excitation or glueball decay rates. 
This reads
\be\label{eq:lateannsimplified}
\frac{dY_{\chi}}{dz}=-\frac{s \langle \sigma_{\rm rea} v\rangle}{H z} Y^2_{\chi} \,,
\ee
where $z=M_\Q/T$, $Y_{\chi}\equiv n_{\chi}/s$ and $s$ is the entropy density of the Universe.
The equilibrium term can be neglected since
$T_R\le\LDC\ll M_\Q$. Assuming a re-annihilation cross section which is constant and velocity independent~\footnote{As explained 
in Appendix~\ref{app:supp}, the last stage of re-annihilation can be effectively described by a constant cross section;
the latter turns out to be also velocity independent in the relevant region of the parameter space of our theories.},
eq.~(\ref{eq:lateannsimplified}) can be easily integrated analytically; 
one obtains (for $T < T_R$)
\be\label{eq:reannY}
Y_{\chi}(T)^{-1}=Y_{\chi}(T_{R})^{-1} + \frac{2}{3} \left(\frac{s \sigma_{\rm rea} v}{H} \right)_{T_R} 
\left[1 - \left(\frac{T}{T_R} \right)^{3/2}\right]\, .
\ee
Late-time annihilation significantly affects the gluequark relic density when the second term in the above equation dominates, \ie roughly when
\be\label{eq:reanncondition}
n_{\chi} \sigma_{\rm rea} v \gg H \qquad \text{at $T =T_R$}\,,
\ee 
in agreement with a naive expectation.
When condition \eqref{eq:reanncondition} is met, any dependence from the previous stages of cosmological evolution, encoded in $Y_\chi(T_R)$, is washed out and the asymptotic value of the relic density is set only by re-annihilation. 
For temperatures $T$ sufficiently smaller than $T_R$ (but higher than a possible subsequent period of dilution, in the case $T_R \sim \LDC$), eq.(\ref{eq:reannY}) can be recast in terms of the gluequark relic density as follows:
\begin{equation}
\label{eq:nchireann}
n_\chi(T) \simeq 1.4 \, \frac{(M_\Q T_R)^{3/2}}{M_\Q M_{\rm Pl}} \frac{g_{\rm SM}(T)^{1/2}}{\sigma_{\rm rea}} \left(\frac{T}{T_R}\right)^3 
\qquad  \text{for \ }  T\ll T_R \, .
\end{equation}
%

\section{Estimate of the Relic density}
\label{sec:relicdensity}

The cosmological evolution of gluequarks is determined by the interplay of the mechanisms described in the previous section
and depends on the two fundamental parameters $\MQ$ and $\LDC$.  For each point in the plane $(\LDC,\MQ)$ one can thus in principle
reconstruct the thermal history of the Universe and compute the DM abundance $\Omega_{\chi}$. 
In this section we will sketch the different possible thermal histories and give an estimate for $\Omega_{\chi}$. 
As a reference model we consider the minimal module with a triplet of $\SU(2)_L$ (see Table~\ref{tab:reps}). We will assume the theory to be outside its conformal window, so that the regime of light dark quarks is well defined. 
We will discuss at the end how the picture changes for different SM representations and when the theory 
is in the conformal window or is not asymptotically free.

\begin{figure}[t]
\centering
\includegraphics[width=.47\textwidth]{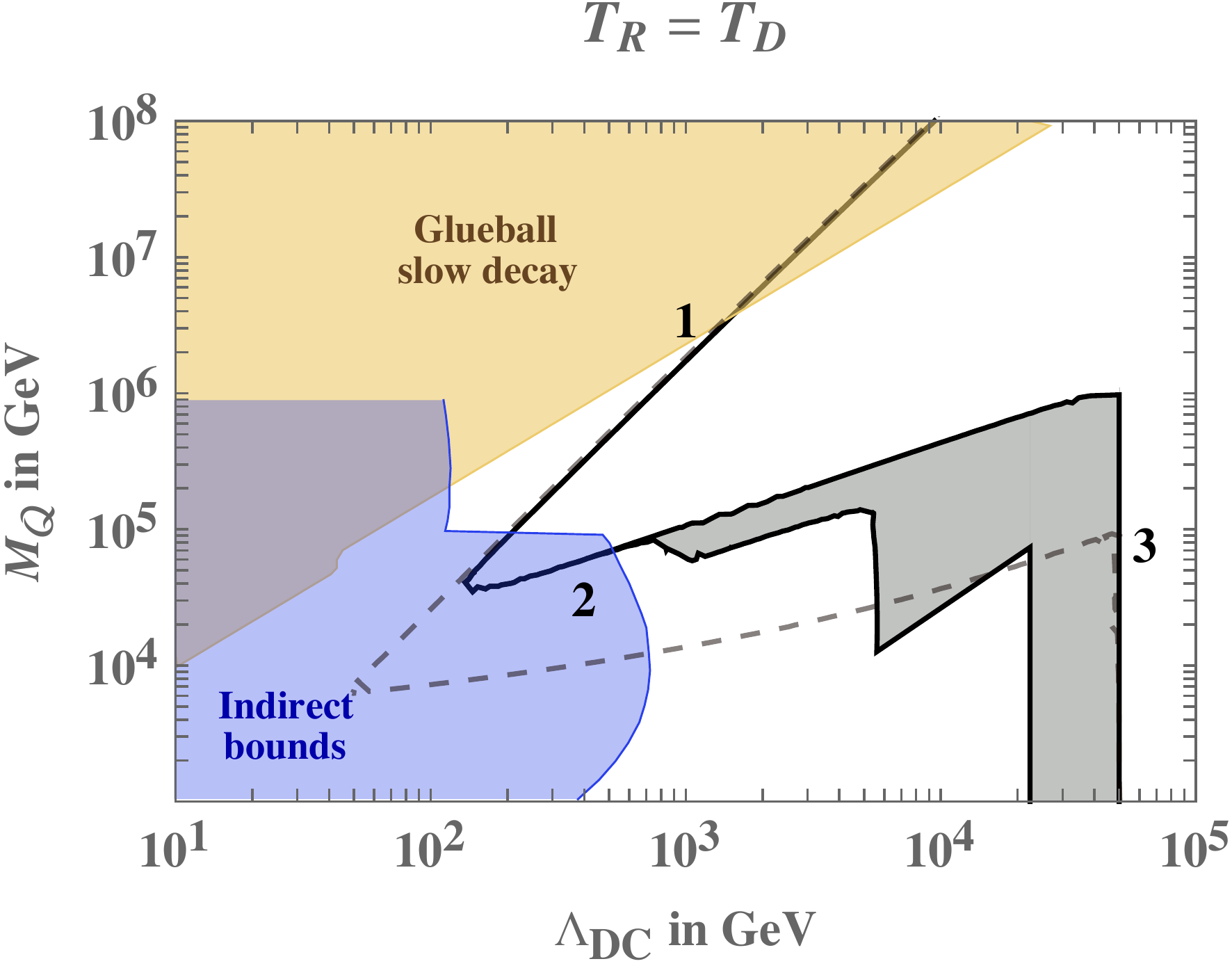}\quad\quad
\includegraphics[width=.47\textwidth]{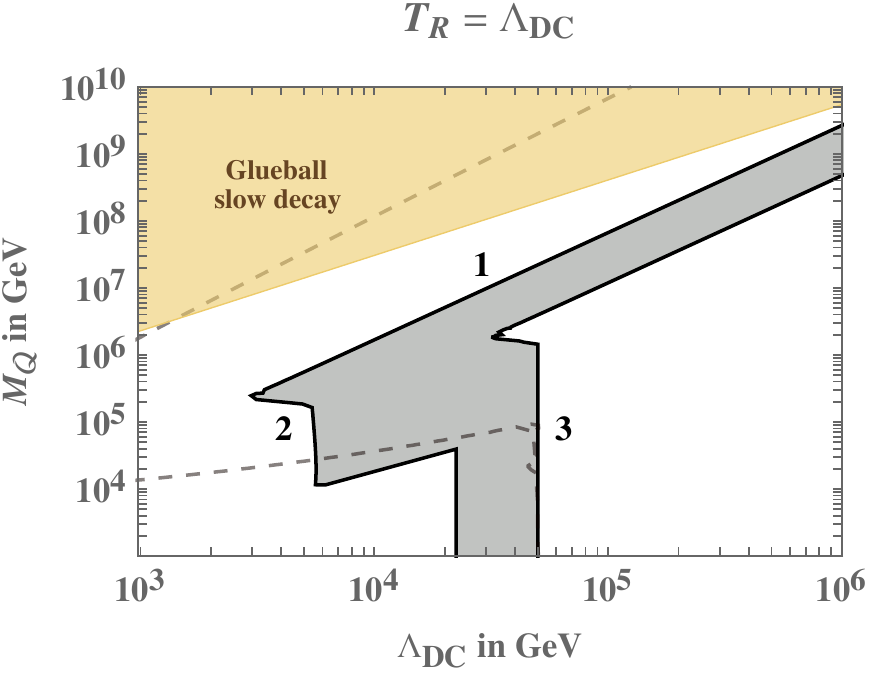}
\caption{\emph{On the black solid line $\Omega_\chi=\Omega_{\rm DM}$. We also report (dashed line) the isocurve $\Omega_\chi=\Omega_{\rm DM}$ for the case where re-annihilation is not considered. The numbers indicate the three thermal histories described in the text. In the yellow region the glueballs are either stable or have a lifetime bigger than $1\s$. In the first case they will over-close the Universe while in the latter they will spoil BBN, both cases are therefore forbidden. The blue region is ruled out by indirect searches, namely modifications of the CMB power spectrum, 21-cm line observables and indirect detection (see section~\ref{subsec:indirect}).}}
\label{fig:relic}
\end{figure}

We will try to quantify the large uncertainties that arise in the determination of the cosmological evolution and of the relic density
as a consequence of the non-perturbative nature of the processes involved.
As anticipated in section~\ref{subsec:reann}, one of the largest uncertainties comes from the identification of the re-annihilation 
temperature~$T_R$.
We will consider the two previously discussed benchmarks: $T_R=T_D$, the most plausible one according to our estimates, and $T_R=T_{dc}$. We reconstruct for each of them the different possible 
cosmological evolutions obtained by varying $\MQ$ and $\LDC$. Our estimate of the DM abundance for both benchmarks is reported 
in Fig.~\ref{fig:relic}, where we show the isocurve $\Omega_{\chi} h^2= 0.119$ reproducing the experimentally observed density. 

Let us consider first the case $T_R = T_D$. There are three possible thermal histories that can be realized (they are correspondingly
indicated in the left plot of Fig.~\ref{fig:relic}):
\begin{enumerate}
\item For very large $\MQ/\LDC$ the Universe undergoes a first perturbative freeze-out at $T_{\rm f.o.}\sim \MQ/25$, then
dark confinement occurs at $T\sim \LDC$ followed by an epoch of dilution between $T_i$ and $T_f=T_D = T_R$~\footnote{Here we are
implicitly assuming that the re-heating temperature at the end of inflation is larger than $\MQ$, so that the number density of dark quarks 
after the perturbative freeze-out is thermal.}.
Glueballs decay at $T \lesssim T_D$, and the number density of gluequarks is too small, as a consequence of the dilution,
to ignite a phase of non-perturbative re-annihilation. The DM density is therefore given by 
\begin{equation}
\label{eq:OmegaA}
\Omega_{\rm DM}\simeq \frac{n(T_{\rm f.o.})  M_\chi}{\rho_{\rm crit}}\left(\frac{T_0}{T_{\rm f.o.}}\right)^3 \mathcal{F}\,,
\end{equation}
where the number density at freeze-out is estimated by solving the Boltzmann equations numerically and approximately given by eq.~(\ref{eq:freezeout}).
By using the dilution factor reported in eq.~(\ref{eq:dilution2}), setting  $M_\chi = M_\Q$, $T_{\rm f.o.} = \MQ/25$, and $T_{dc} = \LDC$ as indicated by lattice studies~\cite{hep-lat/9602007}, one obtains
\begin{equation}
\Omega_{\rm DM}h^2 \sim 0.1 \left(\frac{0.1}{\adc}\right)^2\left(\frac{\LDC}{\TeV}\right)^{3/2}\left(\frac {100\,\LDC}{M_\Q}\right)^2\, ,
\end{equation}
which describes well the slope of the upper part of the relic density isocurve in the left panel of Fig.~\ref{fig:relic}.
Because of the extreme dilution happening during the early epoch of matter domination, the experimental DM abundance is reproduced
in this case for very large DM masses, of order of hundreds of TeV or more, above the naive unitarity bound.
%
\item For smaller values of $\MQ/\LDC$ (but still with $\MQ/\LDC \gtrsim 25$), the dilution between $T_{\rm f.o.}$ and $T_{D}$ is not enough to prevent 
re-annihilation (\ie condition \eqref{eq:reanncondition} is met). The latter thus occurs at $T\simeq T_D$, washing
out any dependence of $\Omega_{\chi}$ from the previous stages 
of cosmological evolution; the corresponding DM relic density is 
\be\label{eq:branchB1}
\Omega_{\rm DM}\simeq\frac{n_\chi(\bar T) M_\chi}{\rho_{\rm crit}}\left(\frac{T_0}{\bar T}\right)^3 \qquad \text{for $T_0 <\bar T\ll T_R$.}
\ee
The first factor corresponds to the gluequark energy density at the end of the re-annihilation (given by eq.\eqref{eq:nchireann}), and the second one encodes 
the standard dilution due to the Universe expansion. We evaluate the re-annihilation cross section by using the semiclassical model described
in Appendix~\ref{sec:npxsec}; this gives
\begin{equation}
\label{eq:modelestimate}
\sigma_{\rm rea}^{model} = 
\displaystyle \frac{4\pi}{\LDC^2} \left[ \varepsilon_\Phi(\LDC,\MQ,T_R) + \alpha_2 \, \varepsilon_V(\LDC,\MQ,T_R) \right]\, .
\end{equation}
The parameters $\varepsilon_\Phi$ and $\varepsilon_V$ are smaller than 1 and encode the suppression from energy and angular 
momentum conservation respectively for the recombination processes 
$\chi \chi \to QQ^* + \Phi$ and $\chi \chi \to QQ^* + V$.
While eq.\eqref{eq:modelestimate} is the result of a rather sophisticated analysis of the re-annihilation dynamics
and represents our best estimate for $\sigma_{\rm rea}$, it is subject to large theoretical uncertainties, as discussed in Appendix~\ref{sec:npxsec}.
We thus also consider the extreme situation where the re-annihilation cross section is always large and
saturated by its geometric value
\begin{equation}
\label{eq:geometric}
\sigma_{\rm rea}^{geo} = \frac{4\pi}{\LDC^2}\, .
\end{equation}
Varying $\sigma_{\rm rea}$ between the values in eqs.\eqref{eq:modelestimate} and \eqref{eq:geometric} will quantify the uncertainty 
on $n_\chi(\bar T)$.
By using eq.\eqref{eq:nchireann}  and setting $M_\chi = M_\Q$ and $T_R = T_D\sim \sqrt{\MPL \Gamma_\Phi}$, eq.~\eqref{eq:branchB1} takes the form:
\be\label{eq:branchB2}
\Omega_{\rm DM}h^2 \sim 0.1
\left(\frac{\LDC}{\GeV}\right)^{11/4}\left(\frac {M_\Q}{1000\,\LDC}\right)^{15/2} \frac{4 \pi/\LDC^2}{\sigma_{\rm rea}(\MQ,\LDC)} \,.
\ee
This formula describes the intermediate part of the isocurve in the left plot of Fig.~\ref{fig:relic}.
Initially (\ie for $150\,\text{GeV}\lesssim \LDC\lesssim 800\,\text{GeV}$) the re-annihilation is dominated by the process $\chi \chi \to QQ^* + \Phi$ 
and $\varepsilon_\Phi \simeq 1$; in this case the last factor in \eqref{eq:branchB2} can be well approximated with 1 (the 
electroweak contribution to $\sigma_{\rm rea}^{model}$ is small) 
and the estimated uncertainty on the gluequark relic density is negligible. 
For larger $\LDC$ 
re-annihilation into $QQ^*$ plus a glueball becomes kinematically forbidden in our semiclassical model,
and $\varepsilon_\Phi$ quickly drops to zero (see Appendix~\ref{sec:npxsec}). 
In this region $\varepsilon_V \simeq 1/10$ and 
varying $\sigma_{\rm rea}$ between $\sigma_{\rm rea}^{model}$ and $\sigma_{\rm rea}^{geo}$ spans the gray region. The extension of the
latter quantifies the uncertainty of our estimate of the relic density.
%
\item When $\MQ/\LDC \leq 25$, the perturbative freeze-out does not take place. If $\MQ$ is bigger than $\LDC$, then the Universe undergoes 
a first epoch of annihilation of dark quarks for $T \gtrsim \MQ$, followed after confinement by the annihilation of gluequarks, until
thermal freeze-out of these latter occurs at $T \simeq M_\chi/25$. If $\MQ < \LDC$, on the other hand, the theory is in its light quark regime
and the only epoch of annihilation is that of gluequarks after dark confinement, 
again ending with a freeze-out at $T \simeq M_\chi/25$.
Afterwards $n_\chi$ is diluted by the Universe expansion without any enhancement from the decay of glueballs 
(these are too short lived to give an early stage of matter domination). The expression for the DM relic density is formally the same as
in eq.\eqref{eq:OmegaA} with $\mathcal{F}=1$. Setting $T_{\rm f.o.} = M_\chi/25$, one obtains
\be\label{eq:OmegaC}
\Omega_{\rm DM}h^2\approx 0.1\, \frac{4\pi/\LDC^2}{\sigma_{\rm ann}} \left(\frac{\LDC}{100 {\rm TeV}}\right)^2\,.
\ee

For $1 \lesssim \MQ/\LDC \lesssim 25$ the non-perturbative annihilation of gluequarks proceeds through the same
recombination processes of eq.\eqref{eq:recomb}. According to the model of Appendix~\ref{sec:npxsec}, only the final state with a vector boson
is kinematically allowed, and $\varepsilon_V \simeq \alpha_2/10$. This implies $\sigma_{\rm ann} \simeq (\alpha_2/10)\, 4\pi/\LDC^2$, 
so that the DM relic density turns out to be independent of $\MQ$.
If instead the re-annihilation cross section is estimated by eq.\eqref{eq:geometric}, then by continuity with the previous cosmological evolution
one must take $\sigma_{\rm ann} \simeq  4\pi/\LDC^2$, which also corresponds to a relic density independent of $\MQ$. 
Varying $\sigma_{\rm ann}$ between these two values gives the largest vertical portion of the gray region in the left plot of Fig.~\ref{fig:relic}.

As soon as one enters the light quark regime, $\MQ < \LDC$, the annihilation of gluequarks proceeds through the direct annihilation
of their constituents (the theory at $\MQ$ is non-perturbative) with a cross section $\sigma_{\rm ann} = 4\pi c/\LDC^2$, where $c$ is an order 1
coefficient. We vary $1/5 < c < 1$ to quantify the uncertainty in this last non-perturbative process. We thus obtain the narrower vertical
portion of the gray region in the left plot of Fig.~\ref{fig:relic}, which extends down to arbitrarily small $\MQ$. The observed relic density 
in this regime is reproduced for $\LDC \simeq 50\,$TeV, similarly to the light quark regime in baryonic DM models~\cite{1503.08749}.
\end{enumerate}

Let us turn to the case $T_R = T_{dc} = \LDC$. As for $T_R = T_D$ one can identify three possible thermal histories (correspondingly indicated in the
right panel of Fig.~\ref{fig:relic}):
\begin{enumerate}
%
\item For $\MQ/\LDC \gg 25$ the Universe goes first through a perturbative freeze-out of dark quarks at $T_\text{f.o.} \simeq \MQ/25$,
then re-annihilation occurs right after confinement for $T\simeq \LDC$. Finally, dilution takes place between $T_i$ and the temperature of the
glueball decay $T_D$.  The DM relic density is given by the expression in eq.\eqref{eq:branchB1} times the dilution factor $\cal F$. Numerically one has
\be 
\Omega_{\rm DM} h^2 \simeq 5 \cdot 10^{-2} \,\dfrac{\LDC^{4}}{\xi\,\MQ^{5/2}}  \frac{4\pi/\LDC^2}{\sigma_{\rm rea}(\MQ,\LDC)}\,.
\ee
In this case, our semiclassical model estimates $\varepsilon_\Phi \simeq 1/100$ throughout the parameter space of interest.
By varying $\sigma_{\rm rea}$ between $\sigma_{\rm rea}^{model}$ and $\sigma_{\rm rea}^{geo}$ we thus obtain the upper portion of the gray region 
in the right plot of Fig.~\ref{fig:relic}.
%
\item For smaller $\MQ/\LDC$ (but still with $\MQ/\LDC > 25$), the glueballs are too short lived to ignite the dilution, and the DM relic density
is given by eq.\eqref{eq:branchB1}. Setting $T_R =\LDC$ one obtains
\begin{equation}
\Omega_{\rm DM} h^2\simeq 10^{-10}\,\frac{\LDC M_\Q}{\GeV^2}\left(\frac{M_\Q}{\LDC}\right)^{1/2}
\frac{4\pi/\LDC^2}{\sigma_{\rm rea}(\MQ,\LDC)}\,.
\end{equation}
%
\item When $\MQ/\LDC < 25$ the cosmological evolution of the Universe is the same as thermal history 3 in the case $T_R =T_D$.
The DM relic density is given by eq.\eqref{eq:OmegaC}, corresponding to the vertical gray regions of the right plot of Fig.~\ref{fig:relic}.
\end{enumerate}

The plots of Fig.~\ref{fig:relic} graphically summarize our estimate of the DM relic density including the uncertainty from the
value of $T_R$ (left vs right panel), and from the value of the cross sections for gluequark re-annihilation and annihilation 
in the light-quark regime (gray region). Reducing substantially the uncertainty on the re-annihilation process (both the cross section and the
value of $T_R$) is not simple and would require a dedicated and in-depth study of the recombination and de-excitation rates, and an extensive study of the system of Boltzmann equations, which is
beyond the scope of the present work. An improved precision in the context of our semiclassical model, on the other hand,
could be obtained from a more accurate knowledge of the spectrum of states in the strong sector, in particular of the masses
of the glueball and gluequark; this can be obtained through dedicated
lattice simulations. Notice also that the plots of Fig.~\ref{fig:relic} have been obtained by assuming a dark color gauge group $SU(3)$, 
for which the confinement temperature $T_{dc}$  and the non-perturbative matrix element relevant for the glueball decay rate are known 
from lattices studies.
Extending our results to other dark gauge groups would in general require to determine these inputs with dedicated simulations, in absence
of which there would be further theoretical uncertainties (both in the estimate of the DM relic density, through the 
expression of the dilution factor in eq.~\eqref{eq:dilution2}, and in the exclusion region from the glueball lifetime).

As a last remark we notice that the qualitative picture derived in this section is largely independent of the details of the specific model. However, the quantitative results can change significantly in models with Yukawa couplings, where the glueball lifetime is much shorter. 
In particular, the exclusion region from the glueball lifetime moves further up left and
branch 1, where dilution occurs, becomes vertical (so that the relic density is uniquely fixed in terms of $\LDC$).
Finally, models that, in the limit of zero quark masses, are infrared free or in the conformal window are constrained to be in the regime $\MQ > \LDC$. 


\section{Phenomenology and Experimental Constraints}
\label{sec:bounds}

In this section we outline the main phenomenological signatures for collider physics and cosmology of the models with gluequark DM.
In general, the phenomenology has analogies to the one of baryonic DM studied in Refs.~\cite{1503.08749,1707.05380}. 
Given the large gluequark masses needed to reproduce the DM relic density both in the light and heavy quark regimes, 
searches at colliders are not promising, whereas cosmological observations provide interesting bounds.


\subsection{Collider searches}

The dark sector has a rich spectrum of states which, in principle, one would like to study at colliders.

The lightest states in the spectrum, with mass given by eq.\eqref{eq:pionmass}, are the NGBs from the $\SU(N_F)\to \SO(N_F)$ global symmetry 
breaking in the light quark regime.
In the case of the $V$ model, the five NGBs form a multiplet with weak isospin 2, and one expects $m_\varphi\gtrsim \LDC/5$.
The phenomenology of a quintuplet of NGBs was studied recently in Ref.~\cite{Barducci:2018yer}. These states are pair produced at hadron 
collider in Drell-Yan processes through their electroweak interactions, and decay to pairs of electroweak gauge bosons through the anomalous
coupling
\begin{equation}
\frac {2 \NDC}{\sqrt{3}} \frac{\alpha_2^2}{4\pi}\frac{\varphi_{ab}}{f} W_{\mu\nu}^a\tilde{W}^{b\,\mu\nu}\, .
\end{equation}
A promising discovery channel studied by Ref.~\cite{Barducci:2018yer} is $pp\to \varphi^0 \varphi^\pm \to 3\gamma W^\pm$;
the doubly charged states decay into same-sign $W$ pairs and are somewhat more challenging to see experimentally.
The LHC has an exclusion reach up to TeV masses, while a $100\,$TeV collider would test the light quark scenario approximately up to 5 TeV.
In this regime colliders could start probing the thermal region. 

The lightest states in the heavy quark regime are the glueballs. They couple to the SM only through higher-dimensional operators, and
are rather elusive at colliders. In models without Yukawa couplings, where interactions with the SM occur through dimension-8
operators, the production cross section via vector boson fusion (VBF) or in association with a SM vector boson is too small to observe a signal
in current or future colliders (for example, the VBF cross section at a $100\,$TeV collider is of order $\sigma(pp\to \Phi + jj) \lesssim 10^{-9}\,$fb for 
$\MQ/\LDC =10$ and $M_\Phi > 500\,$GeV).
In models with Yukawa couplings the glueballs mix with the Higgs boson and production via gluon-fusion becomes also possible. 
While this leads to larger cross sections, the corresponding rate is too small to see a signal at the LHC and even high-intensity experiments
like SHiP can only probe light glueballs in a region of parameter space that is already excluded by EW precision tests~\cite{1707.05380}.

Mesons can give interesting signatures in both light and heavy quark regimes.
Bound states made of a pair of dark quarks, $QQ$, can be singly produced through their EW interactions. 
While the production of spin-0 mesons is suppressed since they couple to pairs of EW gauge bosons,
spin-1 resonances mix with the SM gauge bosons of equal quantum numbers and can be produced via Drell-Yan processes.
In the narrow width approximation the cross section for resonant production can be conveniently written in terms of the decay partial widths as
\begin{equation}
 \sigma(pp\to QQ) =\frac{(2 J_{QQ}+1) D_{QQ}}{sM_{QQ}} 
\sum_{\mathcal P} C_{{\mathcal{P P}}}  \Gamma(QQ\to \mathcal {P P}) \,,
\label{eq:singleproduction}
\end{equation}
where $D_{QQ}$ is the dimension of the representation, $J_{QQ}$ the spin, $\mathcal P$ the parton producing the resonance
and $C_{{\mathcal{P P}}}$ are the dimension-less parton luminosities.

\begin{figure}[t]
\begin{center}
\includegraphics[width=0.475\textwidth]{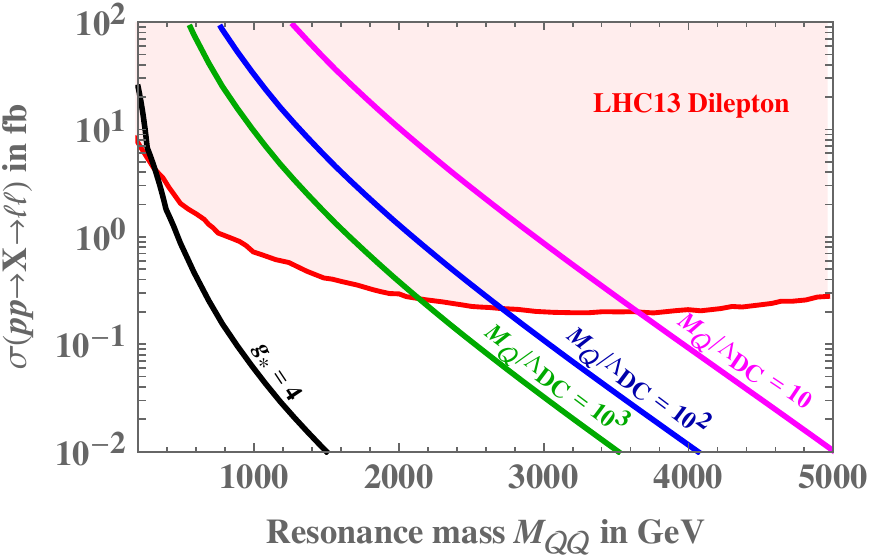}\qquad
\includegraphics[width=0.45\textwidth]{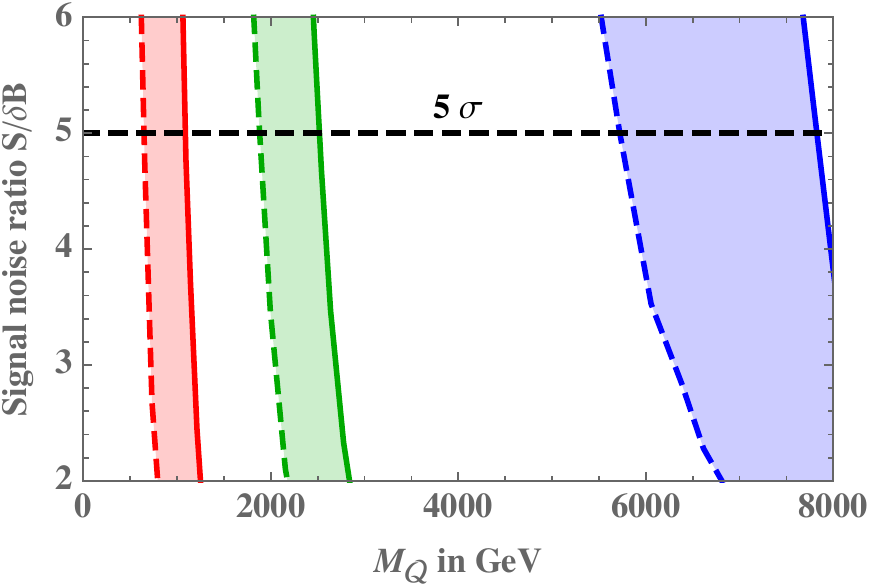}
\caption{\label{fig:collbound}\em {\bf Left:} ATLAS bounds on the cross section for the direct production of a \mbox{spin-1} $QQ$ resonance decaying into muons and electrons~\cite{ATLAS:2017wce}. {\bf Right:} Estimated reach on gluequark pair production obtained by recasting the 
limits of Ref.~\cite{1805.00015} from disappearing tracks searches
at the HL-LHC (red), the HE-LHC (green) and a $100\,\TeV$ collider (blue). The solid (dashed) lines assume a $20\%$ ($500\%$) uncertainty on the background estimate.}.
\end{center}
\end{figure}

In the heavy quark regime the $QQ$ bound state is perturbative and its decay width can be computed by modelling its potential
with a Coulomb plus a linear term. For $\adc M_\Q >  \LDC$ the decay width of the lowest-lying s-wave bound states scales as
\begin{equation}
\label{eq:pertwidthestimate}
\Gamma(QQ\to \mathcal {P P})\sim D_{QQ} \alpha_{SM}^2 \adc^3 M_\Q\, ,
\end{equation}
where $\adc^3$  originates from the non-relativistic Coulombian wave-function.
When $\adc M_\Q<  \LDC$, the effect of the linear term in the potential becomes important and eq.\eqref{eq:pertwidthestimate} gets 
modified; since confinement enhances the value of the wave function at the origin, the width becomes larger in this regime. 
Using the Coulombian approximation thus provides conservative bounds.
Explicit formulas for the rates are found in \cite{1707.05380}. For example, 
in the $V$ model the decay width of the s-wave spin-1 $QQ$ resonance (isospin 1 in light of the Majorana nature of $V$) 
into a left-handed fermion doublet is
 \be\label{eq:rates1ff}
\Gamma^{(n)} \left(QQ^{J=1}_{I=1}\to f\bar f\right)=(N_{\rm DC}^5-N_{\rm DC}^3) \frac{\alpha^2_2\adc^3}{24 n^3}  M_\Q\, ,
\ee
where $n$ refers to the radial quantum number. The tiny energy splitting between levels is irrelevant at colliders and the total rate is 
dominated by the lowest-lying Coulombian ones. 
The branching ratio into pairs of leptons  is about $7\%$ and the strongest bounds currently arise from searches 
of spin-1 resonances at the LHC decaying into electrons and muons. We show the limits in the left plot of Fig.~\ref{fig:collbound}
and find that the LHC excludes masses up to $2-3.5\,$TeV depending on the ratio $\MQ/\LDC$ (or equivalently on the value of $\adc(\MQ)$). 

In the light quark regime the lightest spin-1 state is the $\rho$ meson with mass $M_\rho\sim~\LDC$.  The widths scale as \cite{Contino:2006nn}
\begin{equation}
\begin{split}
\Gamma(\rho \to \varphi\varphi) & \sim \frac{g_*^2}{8\pi} M_{\rho} \\
\Gamma(\rho \to f\bar f) & \sim \alpha_{SM}^2 \left( \frac{4\pi}{g_*^2} \right)M_{\rho} \, ,
\end{split}
\end{equation}
where $g_*$ characterizes the interaction strength among bound states. 
For moderately large~$g_*$, as suggested by large-$N_{DC}$ counting $g_*\sim 4\pi/\sqrt{N_{DC}}$, the decay into light NGBs dominates while final states with leptons are suppressed.
It thus follows a weaker bound than in the heavy quark regime, as illustrated in Fig.~\ref{fig:collbound} for $g_*=4$.

Gluequarks can also be pair produced at colliders through their EW interactions.
In the heavy quark regime the energy threshold is much higher than the confinement scale and quarks are produced in free pairs.
Because dark quarks are in the adjoint representation of dark color, when they get separated by a distance of $O(1/\LDC)$ 
they hadronize producing color singlets that fly through the detector. On the contrary, dark quarks in the fundamental representation would not be able 
to escape, leading to quirks/hidden valley phenomenology~\cite{Kang:2008ea,hep-ph/0604261,1707.05380}.
The phenomenology of the open production is then identical to the one of an elementary electroweak multiplet except that the cross-section 
is enhanced by the multiplicity of the dark color adjoint representation, i.e. $N_{\rm DC}^2-1$ for $\SU(N_{\rm DC})$. 
Such enhancement factor is not present for gluequark pair production near threshold in the light quark regime.
In general, an electroweak triplet can be searched for in monojet and monophoton signals or disappearing tracks, the latter being more constraining. We derived the reach of the high-luminosity LHC (HL-LHC), the high-energy LHC (HE-LHC) and the proposed $100\,$TeV collider by recasting the results of Ref.~\cite{1805.00015} for the $V$ model in the heavy quark regime, see the right plot of Fig.~\ref{fig:collbound}. We find that the HL-LHC could discover gluequark triplets up to $\sim 600\,$GeV while a $100\,$TeV collider could reach $\sim 7\,$TeV.  Such bounds are typically weaker than the ones from the production of $QQ$ spin-1 resonances decaying to leptons.


\subsection{DM Direct Detection}

From the point of view of DM direct detection experiments, where the momentum exchanged is less than 100 KeV, the gluequark behaves 
as an elementary particle with the same electroweak quantum numbers as the  constituent quark.  The main difference from elementary candidates with same quantum numbers is that the relic abundance is not controlled by the electroweak interaction, leading to a different thermal region.

For a triplet of $\SU(2)$ the spin-independent cross-section is $\sigma_{\rm SI}= 1.0 \times 10^{-45}\,$cm$^2$, which is below the 
neutrino floor for masses $M_\chi> 15 $ TeV. 
For an $\SU(2)$ doublet tree-level $Z$-mediated interactions induce a spin-independent cross section on nucleons $\sigma_{\rm SI}\approx 10^{-40}\cm^2$, which is ruled out for $M_\chi\lesssim10^8\GeV$ \cite{1805.12562}. Dark quark masses large enough to make the doublet model viable can be obtained only in the scenario where $T_R=\LDC$, while the scenario with $T_R=T_D$ is ruled out (see Appendix~\ref{app:LL} for more details).


\subsection{DM Annihilation and Decay}
\label{subsec:indirect}
After freeze-out, decays or residual annihilations of gluequarks can give rise to indirect detection signals.

In the region of parameter space relevant for our purposes, annihilation processes set constraints on theories in the heavy quark regime, 
and we thus focus on that case to analyse them.
As discussed in section \ref{subsec:reann} (and more extensively in Appendix~\ref{app:supp}), the annihilation can be either direct ($\chi \chi\to n\Phi/{\rm SM}$) or mediated by
the formation of a $\Q\Q^*$ bound state that subsequently decays ($\chi\chi\to \Q\Q^*\to n\Phi/{\rm SM}$). In the former case the annihilation cross section is perturbative (see eq.\eqref{eq:annDCSM}) and, given the 
relatively high mass of the gluequark, it does not lead to any interesting indirect detection signatures. 
The latter case, instead, because of the enhanced annihilation cross section, could lead to an interesting phenomenology and it is worth further study. 

As discussed in section \ref{subsec:reann} and in Appendix~\ref{app:supp}, for angular momenta $\ell\gg1$ the recombination cross section is of order  $\sigma_{\rm rea}\simeq \varepsilon \,4\pi\//\LDC^2$. However, given the small velocities relevant for indirect searches ($v_{\rm rel}\sim 10^{-6} \sqrt{{\rm TeV}/M_\Q}$ at the CMB epoch and $v_{\rm rel}\sim 10^{-3}$ in our galaxy), the angular momentum of the colliding particles $\ell\sim M_\Q v_{\rm rel}/\LDC$ is of order unity in a large region of the parameter space. In this case only s-wave processes take place and $\sigma_{\rm rea} v_{\rm rel}$ rather than $\sigma_{\rm rea}$ is constant. In this regime the value of the cross section is very uncertain, and we chose to estimate it
in terms of two benchmark scenarios (see Appendix \ref{sec:npxsec}):
\begin{equation}
\langle\sigma_{\rm ann} v_{\rm rel}\rangle\sim 
\begin{cases}
\dfrac{1}{\LDC^2}\\[0.4cm]
\pi R_B^2\approx \dfrac{\pi}{(\adc^2 M_\Q^2)} \,.
\end{cases}
\label{sigmageo}
\end{equation}

Once formed, the $\Q\Q^*$ bound states de-excite and in general decay into dark gluons, SM gauge bosons or SM fermions. 
The branching ratios can be derived in terms of the perturbative annihilation cross section of dark quarks
into the corresponding final states, see eq.\eqref{eq:annDCSM}.  
In the region of interest $\adc > \alpha_2$ and one finds
\begin{equation}
{\rm BR}(GG) \sim1+\mathcal{O}\!\left(\frac{\alpha_2}{\adc}\right) ,
\qquad{\rm BR}(WW/\Psi\Psi) \sim \frac {\alpha_2^2}{\adc^2}, \qquad {\rm BR}(ZG) \sim \frac {\alpha_2}{\adc}\,,
\end{equation}
where $G$ denotes a dark gluon.
For the specific case of the $V$ model, the tree-level decay into SM fermions and $ZG$ is forbidden (the $\chi_0$ has vanishing
coupling to the $Z$) and use of eq.\eqref{eq:annDCSM} thus gives
\begin{equation}
\langle \sigma  v_{\rm rel} \rangle_{\chi_0\chi_0 \to WW}\sim \langle\sigma_{\rm ann} v_{\rm rel}\rangle\times \frac 6 {27} \frac{\alpha^2}{\adc^2}\, ,
\end{equation}
where the last factor is from the branching ratio of $QQ$ into $WW$.

Similarly to residual annihilations, decays of the gluequark could give rise to indirect signals. The $\chi^0$ decays mostly to $h\nu$ plus
glueballs in the heavy quark regime, and to $h\nu$ or $h\nu+\varphi$ in the light quark regime (see eqs.\eqref{eq:lifetimeheavy},\eqref{eq:lifetimelight}).  
Both glueballs and NGBs in turn decay into SM particles and ultimately the gluequark decay leads to 
the production of light SM species which can be observed experimentally. 
Bounds can be avoided, on the other hand, if some mechanism is at work that makes the $\chi^0$ absolutely stable or give it a much longer lifetime than the one estimated in 
eq.~\eqref{eq:lifetimeheavy} and \eqref{eq:lifetimelight}.

\begin{figure}[t]
\centering
\includegraphics[width=.47\textwidth]{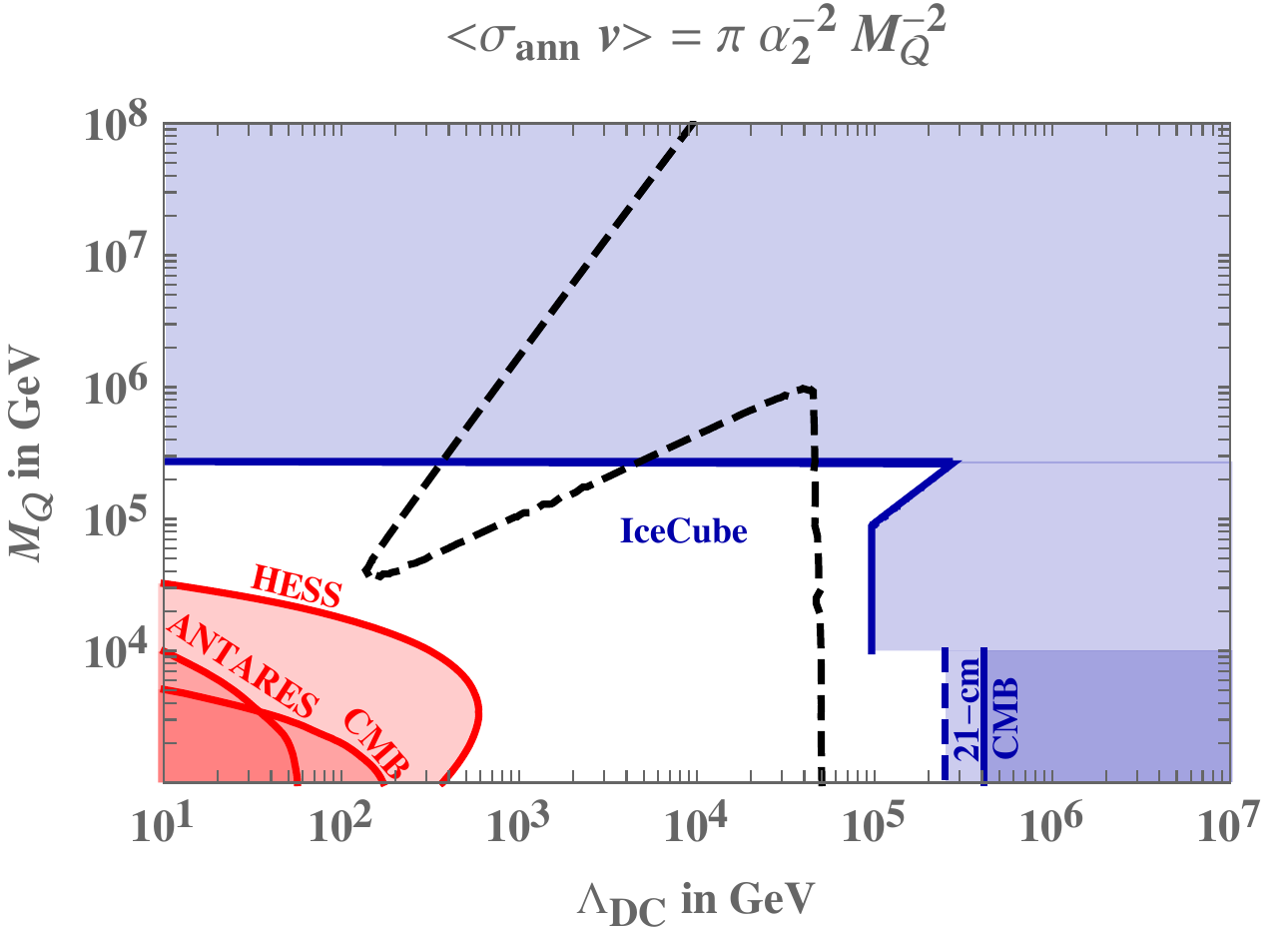}\quad\quad
\includegraphics[width=.47\textwidth]{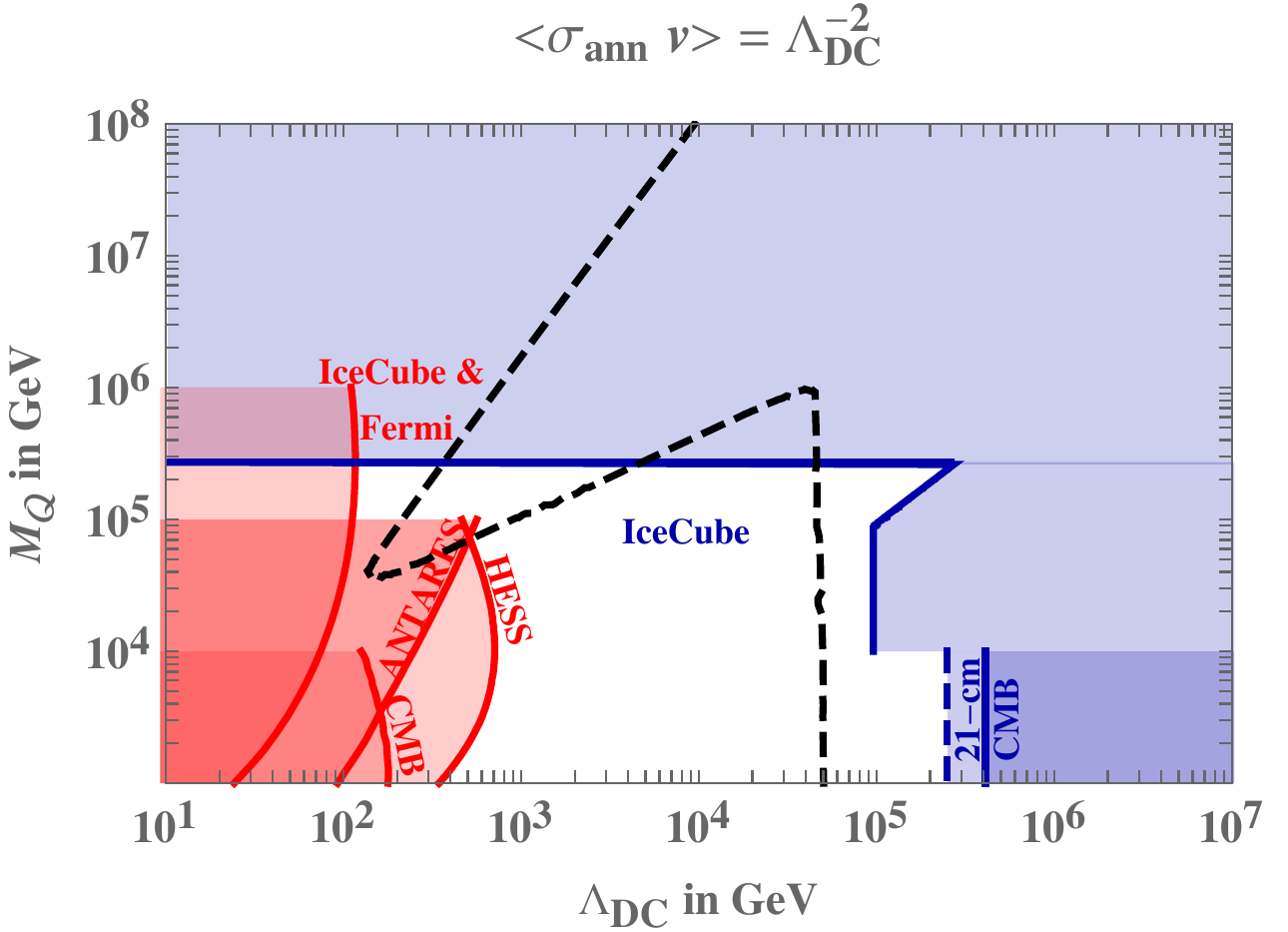}
\caption{\emph{Exclusion limits from experimental searches sensitive to the DM annihilation (red regions) and decay (blue regions).
The limits from DM annihilation respectively in the left and right panels have been obtained by adopting the two benchmarks of eq.~\eqref{sigmageo}, while limits from DM decay were derived by setting $\Lambda_{UV}/g_{UV} = {\bar M}_{\rm Pl} = 2.4\!\times\! 10^{18}\,$GeV 
and $f_\chi = 3\LDC/(4\pi)$ in eqs.~\eqref{eq:lifetimeheavy},\eqref{eq:lifetimelight}.}}
\label{fig:indirect}
\end{figure}

Figure~\ref{fig:indirect} summarizes the constraints in the plane $(\LDC,\MQ)$ that arise from experiments probing DM decay and annihilation.
The red exclusion regions from DM annihilation have been derived for the two benchmark values of $\langle \sigma_\text{ann} v_\text{rel}\rangle$ in eq.~\eqref{sigmageo}, while the blue ones from DM decay were obtained by setting $\Lambda_{UV}/g_{UV} = {\bar M}_{\rm Pl} = 2.4\times 10^{18}\,$GeV and $f_\chi = 3\LDC/(4\pi)$ when evaluating $\tau_{\rm DM}$ from eqs.~\eqref{eq:lifetimeheavy},\eqref{eq:lifetimelight}.
Experimental bounds are given in terms of the DM mass $M_\chi$; in order to translate them into the $(\LDC,\MQ)$ plane we set $M_\chi = \MQ$ in the heavy quark regime and $M_\chi = \LDC$ in the light quark regime. 

The constraints from DM annihilation are characterized by a large theoretical uncertainty, as one can easily see by comparing the 
left and right panels in the figure. Resolving such uncertainty would require a precise determination of the recombination cross section, which does not
seem an easy task in general and is beyond the scope of this work. 
Also the exclusion curve from DM decay has a sizable theoretical uncertainty, which largely comes from the unknown relation between $M_\chi$
and $\LDC$ in the light quark regime (needed to translate the experimental bounds into the $(\LDC,\MQ)$ plane), and from the absence of a calculation of the gluequark decay constant (which controls the size of the
DM decay rate and for which we were only able to give an order-of-magnitude estimate).
In this case, however, dedicated lattice simulations could 
determine these quantities and thus drastically reduce the theoretical error on the blue exclusion curves.

The results of Fig.~\ref{fig:indirect} stem from three classes of experiments, which are discussed in the following.

\subsubsection{Cosmic Rays Experiments}
Given the large gluequark mass needed to reproduce the DM relic density in the heavy quark regime, the strongest indirect detection bounds on DM annihilation
come from the ANTARES neutrino telescope~\cite{1612.04595}, HESS \cite{1607.08142} and the multi-messenger analysis made by the Fermi gamma-ray telescope 
and IceCube~\cite{1206.2595}. The bounds can be roughly summarized as follows~\footnote{Here and in eqs.~\eqref{CMBdecay},\eqref{21cmdecay}
we omit for simplicity the mild dependence that the bounds have on the DM mass. The exclusion curves of Fig.~\ref{fig:indirect} have been obtained by using the 
exact expressions without performing such approximation.}:
\be
\begin{aligned}
\langle \sigma_{\rm ann}v_{\rm rel}\rangle &\lesssim 10^{-7}\GeV^{-2} && \text{(from ANTARES, HESS)}\\
\langle \sigma_{\rm ann}v_{\rm rel}\rangle & \lesssim 10^{-5}\GeV^{-2}&& \text{(from Fermi-ICeCube)}\,.
\end{aligned}
\ee
Indirect searches also place bounds on the lifetime of heavy DM candidates. In the high-mass range, Ice-Cube provides the most
stringent bounds~\cite{1804.03848}. 
For $M_\chi=(10^5 \divisionsymbol 10^7)\GeV$, they are roughly given by
\be
\tau(\chi^0)\gtrsim10^{26} \left(\frac{M_\chi}{100\TeV}\right)^{3/2}\s\,.
\ee

\subsubsection{CMB power spectrum}
The energy released by gluequark annihilations and decays around the epoch of recombination modifies the CMB power spectrum. 
This, similarly to indirect detection experiments, constrains the lifetime and the annihilation rate of the DM. The annihilation cross section is constrained to be smaller 
than \cite{1502.01589}
\be
 \langle \sigma_{\rm ann} v_{\rm rel} \rangle \lesssim 10^{-8}\left(\frac{M_\chi}{100\GeV}\right)
\GeV^{-2}\,,
\ee
while the limits on the DM lifetime are \cite{1610.06933} 
\be
\label{CMBdecay}
\tau\left(\chi^0\right)\gtrsim 10^{24}\s\,.
\ee
These bounds are slightly less stringent than the ones coming from indirect detection, but have the advantage to be free from astrophysical uncertainties. 
They are provided for DM masses up to $10\TeV$, but are expected to be approximately mass-independent for masses above this value \cite{Slatyerprivate}. 
The CMB bounds shown in Fig.~\ref{fig:indirect} have been obtained under this assumption.

\subsubsection{21 cm line}
While CMB is sensitive to sources of energy injection at the epoch of recombination, the cosmic $21$-cm spectrum is sensitive to sources of energy injection during the dark ages. The recent observation of an absorption feature in this spectrum, if confirmed and in agreement with standard cosmology, can be used to put bounds on both the lifetime and the DM annihilation cross section. Conservative limits can be derived by neglecting astrophysical heating sources; the one on annihilation is of order \cite{1803.03629,1803.09739}:
\be
 \langle \sigma_{\rm ann} v_{\rm rel} \rangle\lesssim10^{-5}\left(\frac{M_\chi}{10 \TeV}\right)\GeV^{-2},
\ee
while the one on the DM lifetime is \cite{1803.09390,1803.09739,1803.11169}
\be
\label{21cmdecay}
\tau\left(\chi^0\right)\gtrsim10^{25}\s.
\ee
The latter is independent of astrophysical uncertainties on the distribution of DM.

As for the case of CMB, these bounds are provided up to $M_{\chi}=10\TeV$ and to obtain Fig.~\ref{fig:indirect} we assumed that they are constant at higher masses. 
Differently from the previous case, however, this assumption is not completely justified and further studies are needed to provide solid bounds in the high-mass range~\cite{Slatyerprivate}.

\subsection{Glueball lifetime}

In the region of the parameter space that we consider, $\LDC\gtrsim \rm GeV$, glueballs with lifetime larger than $1\, \rm s$ are excluded by a combination of bounds. 
Cosmologically stable glueballs have a too large relic density and overclose the Universe. Long-lived glueballs, on the other hand, are constrained by BBN observations~\cite{hep-ph/0604251} in the range $ 1 \, {\rm s} < \tau_{\Phi} < 10^{12} \, {\rm s}$ and by observations of the diffuse gamma-ray spectrum~\cite{hep-ph/9610468} in the range $ 10^{12}\, {\rm s} < \tau_{\Phi} < 10^{17} \, {\rm s}$.

These bounds constrain the high-mass region of the $V$ model as shown in Fig.~\ref{fig:relic}. Notice, however, that they could be potentially relaxed if glueballs decay 
through dimension-6 operators (generated for example in models with Yukawa couplings).

\section{Summary}
\label{sec:conclusions}

In this work we continued the systematic study of gauge theories with fermions in real or vector-like representations, initiated
in Ref.~\cite{1503.08749}, where a DM candidate arises as an accidentally stable bound state of the new dynamics.
We considered in detail the gluequark DM candidate, a bound state of adjoint fermions with a cloud of gluons, stable due to dark fermion number.
What makes this scenario special in the context of accidental DM is that the physical size of DM, that controls the low-energy interactions, 
is determined by the dynamical scale of the gauge theory independently of its mass.  
In the heavy quark regime the DM mass and size can be vastly separated  leading to an interesting interplay of elementary and composite dynamics.  In particular, cross sections much larger than the perturbative unitarity bound of elementary particles can arise, modifying the thermal abundance and producing potentially observable signals in indirect detection experiments. Gluequarks display a rich and non-standard cosmological history and could be as heavy as PeV if their abundance is set by thermal freeze-out.

Our estimates show that the observed DM density can be reproduced by gluequarks both in the light and heavy quark regimes.
The mass of the DM is of order $100\,$TeV or larger, which makes the models difficult to be directly tested at present and future colliders.
On the other hand, indirect experiments sensitive to the decay and the annihilation of the DM are a powerful probes of
gluequark theories.
We found that these experiments can already set important limits, excluding part of the curve which reproduces the observed DM density, 
depending on the value of the annihilation cross section and if the naive estimate for the gluequark decay 
rate is assumed (see Fig.~\ref{fig:indirect}). This suggests that gluequark theories in the very heavy quark regime require non-generic UV completions
to ensure the accidental stability of the DM at the level of dimension-6 operators. For example, the dark parity could be gauged in the UV (see Appendix~\ref{app:LL}),
or its violation could be generated only by non-perturbative gravitational effects in a weakly-coupled UV completion. Similar arguments are put
forward also in the context of axion models concerning the quality of the Peccei-Quinn symmetry, see \cite{Redi:2016esr}.
Assuming that an appropriate UV completion exists, gluequark models are interesting examples where the DM density can be generated thermally after inflation by
very heavy particles. This can be contrasted with other scenarios, such as Wimpzillas~\cite{hep-ph/9802238}, where ultra heavy DM candidates are 
never in thermal equilibrium.

The low-energy dynamics and the spectrum of gluequarks are non-perturbative and we were only able to give rough estimates of various effects. In particular,  in the heavy quark regime, the quantitative estimate of the re-annihilation relevant for the thermal relic abundance and indirect detection of DM is highly uncertain, as it depends on the details of the spectrum and on the rates of non-perturbative transitions. A more firm conclusion would require a better knowledge of the recombination cross sections and of the de-excitation rates of bound states, as well as an extensive study of the system of Boltzmann equations.
In the light-quark regime, a non-perturbative calculation of the annihilation cross section would lead to
a sharp prediction of the dynamical scale of the dark sector.
The precise knowledge of the spectrum of gluequarks, mesons and pions would then 
give valuable information for indirect detection and collider studies.

In this work we studied gluequarks as thermal relic candidates and focused on the simplest, minimal theories of Tab.~\ref{tab:reps}.
Investigating non-minimal models would be certainly interesting and important under several aspects. For example, 
SM gauge couplings unify at high energies with less precision in the minimal blocks of Tab.~\ref{tab:reps} than in the SM.
Achieving precision unification thus necessarily requires extending the models to include additional matter with SM quantum numbers.
Furthermore, while the thermal relic abundance hints to a large DM mass, this conclusion can be modified in more general gluequark theories
where the DM is asymmetric (this requires a larger accidental symmetry than dark parity) or where  the DM abundance is determined by the 
decay of unstable heavier states. These theories would have a smaller mass gap and  could be tested at the LHC and at future colliders.
We leave an investigation along these directions to a future work.

\section*{Acknowledgments}
We would like to thank Guido Martinelli, John March-Russell, Filippo Sala, Andrea Tesi, Tracy Slatyer and Hongwan Liu for useful discussions.

\appendix

\section{Dark Quark Annihilation Cross Section}\label{app:ann}

In this section we report the formulas for the annihilation cross section of dark quarks, which are useful 
to study the perturbative freeze-out and DM indirect detection.

Dark quarks can annihilate into dark gluons and into SM final states (mainly $VV$, $Vh$, $hh$ and $\psi\bar\psi$, where $V=W,Z,\gamma$).
These latter contribute significantly to the total cross section in the case of perturbative freeze-out whenever $\MQ/\LDC \gg 1$ and thus
the dark color interaction strength does not exceed much the electroweak one. Final states into SM particles are also expected to be
important for direct detection even though they have a smaller rate compared to DM annihilation into glueballs.

The tree-level annihilation cross-section of dark quarks $\chi_i \chi_j $  in a 
representation $(A_\DC,R_\SM)$ of the dark color and $\SU(2)_L$ groups into massless vectors at low energy reads,
\begin{eqnarray}
&&\langle\sigma v_{\rm rel}\rangle_{ij\to VV} =\frac{A^1_{ij}+A^2_{ij}}{16\pi} \frac{1}{M_\Q^2}\nonumber \\
&&A^1_{ij}\equiv \left[T^a T^a T^b T^b\right]_{ij}\,,~~~~~~~~A^2_{ij}\equiv \left[T^a T^b T^a T^b\right]_{ij}
\end{eqnarray}  
where the generators are written as $T\equiv (g_\DC T_\DC \otimes 1)\oplus (1 \otimes g_\SM T_\SM)$.
Selecting the neutral component in the equation above and averaging over dark color gives the perturbative annihilation 
cross-section of DM today. Averaging over all initial states as required for the thermal freeze-out one finds~\cite{1707.05380},
\begin{equation}
\begin{split}
\langle\sigma v_{\rm rel} \rangle_{\rm ann}=\frac{\pi}{M_\Q^2}\bigg[
& \frac {\adc^2}{d(R_\SM)}\frac {K_1(R_\DC)+K_2(R_\DC)}{g_\chi d(R_\DC)^2 }
+\frac {\alpha_2^2}{d(R_\DC)}\frac {K_1(R_\SM)+K_2(R_\SM)}{g_\chi d(R_\SM)^2 } \\
& +\alpha_\DC\alpha_2\frac {4 C(R_\DC) C(R_\SM)}{g_\chi d(R_\SM) d(R_\DC) }\bigg]\,,
\end{split}
\label{eq:annDCSM}
\end{equation}
where
\begin{equation}
K_1(R)= d(R) C(R)^2,\quad
K_2(R)=K_1(R)-\frac {d(A)C(A)T(R)}2, \quad d(R) = dim(R)\, .
\end{equation}
and $A$ stands for the adjoint representation.
$T(R)$ and $C(R)$ are respectively the Dynkin index and the quadratic Casimir of the representation $R$, and $g_\chi=2(4)$ for real (complex) 
representations.

Dark quarks can also annihilate into final states with SM fermions and Higgs bosons through their SM gauge and Yukawa interactions. These channels have been included in eq. \eqref{eq:pertann}.

\section{Reannihilation}\label{app:supp}

As discussed in section \ref{subsec:reann}, a second stage of annihilation involving 
gluequarks can occur after confinement. The annihilation can proceed in a single step into glueballs or SM vector and Higgs bosons:
\begin{itemize}
\item $\chi + \chi \to n \Phi / nV$: in the heavy quark regime this process has a perturbative cross section; indeed the exchanged momentum in the interaction is of ${\cal O}(\MQ)$ with $\MQ \gg \LDC$, thus the interaction strength is governed by $\gdc(\MQ)$ which is perturbative.
\end{itemize}
\noindent Alternatively, it can take place in two steps, a non-perturbative recombination followed by de-excitation and decay into SM particles: 
\begin{itemize}
\item $\chi + \chi \to \mathcal{Q}\mathcal{Q}^*\dashrightarrow \rm{SM}$: the recombination is two to one and energy conservation implies $M_{\Q\Q^*} > 2M_\chi$, therefore the opposite decay process is always allowed. The matrix element for the inverse decay is non-perturbative and the corresponding rate is expected to be much larger than the rate of the de-excitation process 
$\Q\Q^*\to \Q\Q + n\Phi/nV$.
\item $\chi + \chi \to \Q\Q^* + \Phi / V \dashrightarrow \rm SM $: the recombination takes place with the emission of one 
electroweak gauge boson or, if kinematically allowed, one glueball. Bound states with $M_{\Q\Q^*} < 2M_\chi$  will in general be 
formed which cannot decay back into gluequarks. They can de-excite and decay into SM particles.
The corresponding re-annihilation rate is expected to be non-perturbative and potentially large.
\end{itemize}

\noindent Only the last of the three processes described above can ignite an epoch of re-annihilation. 
The dynamics of re-annihilation is described by a set of coupled Boltzmann equations of the form
\begin{equation}
\label{eq:lateann}
\begin{split}
\frac{dY_{\chi}}{dz} & =
-\frac{s \langle \sigma_{\rm rec} v\rangle}{H z}\left(Y^2_{\chi}-Y_{\chi, eq}^2\,\frac{Y_{\Q\Q^*}\,Y_\Phi}{Y_{\Q\Q^*}^{eq}\,Y_\Phi^{eq}}\right)\,, \\
\frac{dY_{\Q\Q^*}}{dz} & = 
\frac{1}{2} \frac{s \langle \sigma_{\rm rec} v\rangle}{H z}\left(Y^2_{\chi}-Y_{\chi, eq}^2 \frac{Y_{\Q\Q^*} Y_\Phi}{Y_{\Q\Q^*}^{eq}Y_\Phi^{eq}}\right)
-\frac{\Gamma_{\Q\Q^*}}{H z}(Y_{\Q\Q^*}-Y_{\Q\Q^*}^{eq}) \\ 
\frac{dY_{\Phi}}{dz} & =  
\frac{1}{2} \frac{s \langle \sigma_{\rm rec} v\rangle}{H z}\left(Y^2_{\chi}-Y_{\chi, eq}^2 \frac{Y_{\Q\Q^*} Y_\Phi}{Y_{\Q\Q^*}^{eq}Y_\Phi^{eq}}\right)
-\frac{\Gamma_{\Phi}}{H z}(Y_{\Phi}-Y_{\Phi}^{eq})
\,.
\end{split}
\end{equation}
These expressions are simplified in that the actual system of equations involves the number densities of
all possible $\Q\Q^*$ bound states. Furthermore, we have omitted the effect of the recombination 
into EW vector bosons and of the corresponding inverse process. Equation~\eqref{eq:lateann} will be however sufficient for our discussion, 
and the generalization to the full case is straightforward. 

In order to annihilate into SM particles with an unsuppressed rate, an excited $\Q\Q^*$ bound state needs to reach first a state with low angular
momentum. Consequently, re-annihilation is efficient only when the rate of de-excitation 
is larger than the one of dissociation\footnote{In the opposite regime of fast dissociation, and much before the glueball decay, the last term in the second and third lines of
eq.~\eqref{eq:lateann} can be neglected. The solution to the Boltzmann equations is thus given by non-thermal equilibrium values
for the three populations which are close to their initial conditions at dark confinement.}.%

Obtaining a precise estimate of the ratio between the de-excitation and dissociation rates
is difficult because: \textit{i)} the dynamics of these processes is non perturbative and the lifetime depends on the different initial 
and final $\Q\Q^*$ states considered; \textit{ii)}  the rate of the dissociation process initiated by EW vector bosons depends on their energy,
which follows a thermal distribution and thus varies with the temperature. 
The result is that the re-annihilation process can be efficient for some of the $QQ^{\ast}$ states and inefficient for others, and it 
becomes more and more efficient as the temperature decreases. 

This can be effectively described as a non-perturbative re-annihilation process happening with a temperature-dependent cross section that 
saturates to a maximal value when dissociation becomes inefficient for all the bound states.
Since the evolution of the relic density takes place on relatively short time scales, the final abundance after this second freeze-out can be 
approximately characterized by two parameters: the final (maximal) value of the cross section, and the temperature at which this final cross 
section is reached. 
These two quantities will be dubbed respectively as the re-annihilation cross section, $\sigma_{\rm rea}$, and the re-annihilation temperature, 
$T_R$. 

During the last stage of re-annihilation, for sufficiently large $\Gamma_\Phi$ or $\Gamma_{\Q\Q^*}$, the system of equations given in eq.~\eqref{eq:lateann} simplifies. The abundance of gluequarks can be described by a single equation, see eq.~\eqref{eq:reannY}.

\subsection{Estimate of the Re-annihilation Cross Section}
\label{sec:npxsec}

In this section we try to estimate $\sigma_{\rm rea}$
using considerations based on energy and angular momentum conservation and simplified phenomenological models.

First of all, it is useful to determine if (depending on value of the temperature, $\LDC$ and $\MQ$) the recombination 
process takes place in a semiclassical or fully quantum regime. If the De Broglie wavelength of the particle $\lambda =h/p$ is of order 
or larger than the typical interaction range $R\sim 1/\LDC$ the collision is fully quantum mechanical, otherwise a semiclassical picture 
can be adopted.
The condition for a semiclassical behaviour can be recast as $l_{\rm max}  \sim M_\chi v R \gg 1$,
where $l_{\rm max}$ is the maximum angular momentum of the process given the short-range nature of the interaction.
We find that the processes occurring in the very early Universe (at $T=T_R$) are always in the semiclassical regime in the
region of parameter space where the DM experimental density can be reproduced. Recombination processes occurring at the CMB or  at
later times, instead, turn out to be quantum mechanical because of the much lower gluequark velocity.

In the quantum regime, the lowest partial wave is expected to dominate in the low momentum limit $k\to 0$. 
In the case of exothermic reactions~\footnote{Exothermic reactions are those where the particles in the final states are lighter than those
in the initial state.}, as the one considered here, general arguments of scattering theory suggest that the cross section scales 
as $1/k$ for $k\to 0$ if the process can take place in \emph{s}-wave~\cite{Weinberg:1995mt}.
In the non-relativistic limit we thus expect a cross section $\sigma\varpropto1/v$. Since the process is non-perturbative it is not possible to compute this cross section from first principles; furthermore, since two different scales ($\MQ$ and $\LDC$) appear in the problem, it is not clear what is the cross section scaling\footnote{The electroweak process $\chi + \chi \to QQ^* + V$ 
has a close nuclear analogue given by $p+n \to d + \gamma$.
Explicit calculations reproduce the expected $1/v$ velocity dependence \cite{Kaplan:2005es}. The non-perturbative constant in that case can be predicted using elastic nucleon scattering data.}. 
In light of this we adopt two different benchmark scenarios:

\begin{equation}
\langle\sigma_{\rm ann} v_{\rm rel}\rangle\sim 
\left\{
\begin{array}{l}
\dfrac{1}{\LDC^2}\\ \\
\pi R_B^2\approx \dfrac{\pi}{(\adc^2 M_\Q^2)}\, .
\end{array}
\right.
\label{sigmageo}
\end{equation}
In the first one the cross-section is controlled by the size of the gluequark while in the latter is  the size of the {\emph s}-wave ground state which fixes the cross section.

In the semiclassical regime we estimate the re-annihilation cross section 
using a simple dynamical model. We first discuss the process $\chi + \chi \to QQ^* + \Phi$ and 
then analyse the recombination into EW vector bosons.
Our semiclassical model is defined in terms of the following simplified assumptions:
\begin{itemize}
\item The gluequarks are modelled as hard spheres with radius of order $R \sim 1/\LDC$, colliding with impact parameter $b$ and thermal velocity $v$.
\item The interaction is short range, therefore the maximum impact parameter for which an interaction occurs is $b_{\rm max} = 2R \sim 2/\LDC$.
We define a corresponding geometric total cross section
\be
\sigma_{\rm total} = \pi b_{\rm max}^2 = \dfrac{4 \pi}{\LDC^2}\, .
\ee
For thermal velocities, $b_{\rm max}$ can be converted into a maximum angular momentum 
$l_{\rm max} = b_\text{max}M_\chi v \sim 2 (M_\chi/\LDC)\sqrt{3T/M_\chi}$ for the colliding particles.
\item Energy conservation implies that only some bound states can be formed. Among these we identify the states with maximum angular momentum $l_{*}$ allowed by energy conservation and by the short range constraint $l_{*} \leq l_{\rm max}$.
\item Angular momentum conservation implies that only interactions with impact parameter smaller than $b_{*} \approx (l_{*}+1)/(M_{\chi}v)$ can lead to bound state formation~\footnote{The factor $(l_{*}+1)$ takes into account the quantization of $l_{*}$ and ensures that the cross section is not underestimated for small angular momenta.}. The short range interaction constraint then forces $b_{*}\leq b_{\rm max}$. If no bound state is allowed by energy conservation we take $b_*=0$.
\item The recombination cross section is estimated by the geometrical value $\sigma = \pi b_{*}^2$.
\end{itemize}
The model predicts a re-annihilation cross section into glueballs that can be conveniently expressed in terms of a suppression 
factor $\varepsilon_\Phi$ as follows:
\be
\sigma_{\text{rea},\Phi} = \pi b_{*}^2 = \left(\dfrac{b_{*}}{b_{\rm max}}\right)^2 \sigma_{\rm total}\equiv\varepsilon_\Phi\,\sigma_{\rm total}\,,
\ee
where $\varepsilon_\Phi$ is computed to be
\be\label{eq:suppressionfactor}
\varepsilon_\Phi =
\begin{cases}
1 \hspace{100pt} & \mathrm{if}\quad l_{*}> l_{\rm max}-1, \\[0.3cm]
\dfrac{\LDC^{2}}{4}\dfrac{(l_{*}+1)^2}{M^{2}_{\chi}v^{2}}  &  \mathrm{if}\quad l_{*}< l_{\rm max}-1, \\[0.3cm]
0 & \mathrm{if}\quad l_{*} \rm \; does\; not\; exist.
\end{cases}
\ee
Notice that $\varepsilon_\Phi$ is a function of $\MQ$, $\LDC$ and indirectly of the temperature through the value of $l_*$ and $v$.

In order to determine $l_*$ we use the energy balance equation in the center-of-mass frame:
\be\label{eq:energycond}
2 M_\chi+2 K_\chi\geq M_{\Q\Q^*}+M_\Phi\, ,
\ee
where $K_\chi$ is the kinetic energy of the colliding gluequarks. 
The gluequark mass can be written in terms of the quark mass plus a binding energy $B_\chi$:
\be
M_{\chi} = \MQ + B_{\chi}\, .
\ee
Similarly, the mass of the di-quark bound state is decomposed as
\be
M_{\mathcal{Q}\mathcal{Q}^*} = 2\MQ + B_{\mathcal{Q}\mathcal{Q}^*}\, .
\ee
We set the gluequark binding energy to the value computed in QCD lattice simulations of $\SU(3)$ gauge theories: 
$B_{\chi} = 3.5 \, \LDC$~\cite{Bali:2003jq}~\footnote{The bare quark mass and binding energy are renormalization scheme 
dependent. Here we quote the result of reference \cite{Bali:2003jq} valid in the RS scheme which, according to the authors, smoothly converges to the $\overline{MS}$ scheme in the perturbative regime. Since we are interested in
just an order-of-magnitude determination of the relic density, we neglect the scheme dependence of $B_\chi$ in what follows.
We notice however that our numerical estimate of the re-annihilation cross section is rather sensitive to the value of $B_\chi$,
hence the scheme dependence can have a strong impact. We take such theoretical uncertainty effectively into account by considering 
different benchmark scenarios, as explained in section~\ref{subsec:reann}.}.
The binding energy of the $\Q\Q^*$ bound state, $B_{\mathcal{Q}\mathcal{Q}^*}$, is instead approximated by the energy levels of a confining 
model with a Coulomb potential plus a linear term~\cite{Cornell}
\be
V(r) = - \dfrac{\alpha_{\rm eff}}{r} + \sigma r,
\ee
with $\alpha_{\rm eff} = \adc(\MQ)$ and $\sigma = 2.0\,\NDC \LDC^2$.
Therefore, $B_{\mathcal{Q}\mathcal{Q}^*}$ is computed numerically as a function of the principal and orbital quantum numbers of the bound state.
The energy balance of eq.\eqref{eq:energycond} can be rewritten as
\be
\label{eq:energy_balance}
B_{\mathcal{Q}\mathcal{Q}^*} \leq 2B_{\chi} + 2 K_\chi- M_{\Phi}\, ,
\ee
and implies a constraint on the maximal angular momentum $l_*$ (as well as on the principal number).
In general one should also impose the additional condition $M_{QQ^*} < 2M_\chi$, to ensure that the decay of the $QQ^*$ state back
into gluequarks is kinematically forbidden. In terms of binding energies, this condition reads
\begin{equation}
\label{eq:decayforbidden}
B_{\Q\Q^*} < 2 B_\chi\, .
\end{equation}
The average gluequark kinetic energy in eq.~\eqref{eq:energy_balance} is of order of the temperature, which in turn is smaller than $\LDC$.
We set the glueball mass to its value computed on the lattice in $\SU(3)$ Yang-Mills theories, $M_\Phi \simeq 7 \LDC$, and thus
find $2 K_\chi- M_{\Phi} < 0$. As a consequence, the condition \eqref{eq:energy_balance} is always stronger than \eqref{eq:decayforbidden}.

\begin{figure}[t]
\centering
\includegraphics[width=.45\textwidth]{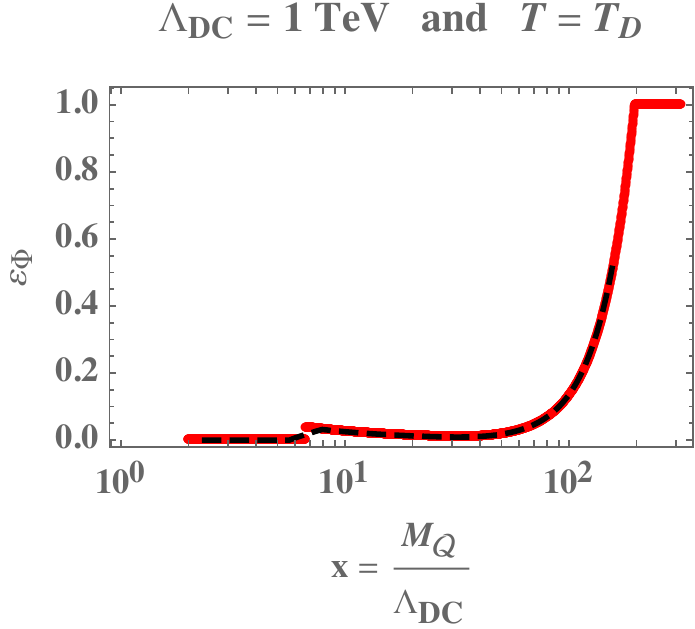}
\quad\quad
\includegraphics[width=.45\textwidth]{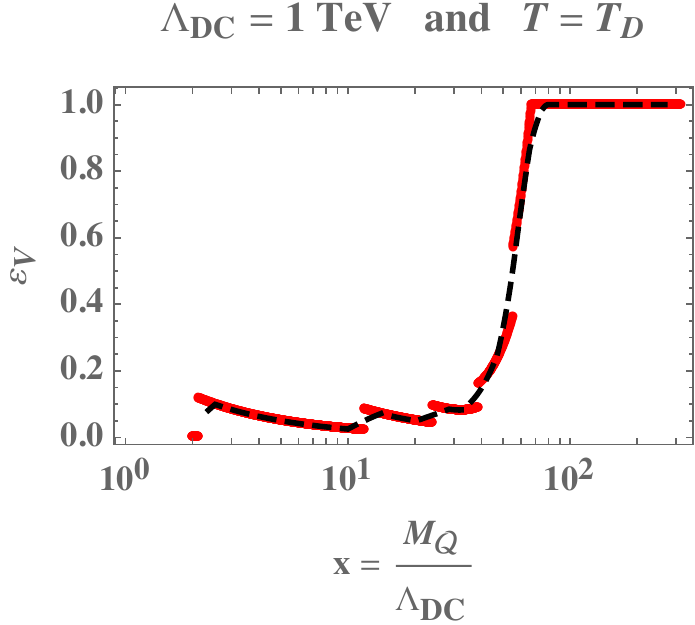}
\caption{\emph{Suppression factors $\varepsilon_\Phi$ (left panel) and $\varepsilon_V$ (right panel) at $T=T_D$ as a function of the ratio 
$M_\mathcal{Q}/\LDC$ for $\LDC=1\TeV$.
In red the results of the numerical computation and in black the interpolation used to compute the relic density. 
Discontinuities in the numerical results are due to the integer nature of $l_*$.}}
\label{fig:suppression}
\end{figure}

Since eq.~\eqref{eq:energy_balance} depends on the gluequark kinetic energy, which we set to $K_\chi = T$ in our numerical computation,
the value of $l_*$ will have a dependence on $T$.
For illustration we show in Fig. \ref{fig:suppression} the value of $\varepsilon_\Phi$ as a function of $\MQ/\LDC$ obtained at $T=T_D$ 
for $\LDC = 1\,$TeV.  Changing $\LDC$ while keeping the temperature fixed leads to very small variations of $\varepsilon_\Phi$.
For $T=\LDC$, on the other hand, $\varepsilon_\Phi$ turns out to be small and of order of a few$\times 10^{-2}$ in the region of interest
($100\,\text{GeV} < \LDC < 10\,\text{TeV}$ and $1 < \MQ/\LDC < 100$).
%
%

In the case of the recombination with the emission of a vector boson, $\chi + \chi \to QQ^* + V$, we expect the re-annihilation cross section 
to be suppressed  by at least a factor $\alpha_2$. Clearly, this process becomes relevant only when the recombination with glueball emission 
is strongly suppressed or forbidden  for kinematic reasons. 
The transitions $\chi + \chi \to QQ^* + V$ that are relevant for re-annihilation are those where 
the $QQ^*$ is sufficiently light so that it cannot decay back in two $\chi$'s. Such bound states satisfy the condition \eqref{eq:decayforbidden},
which requires the kinetic energy of the emitted vector boson to be larger than the sum of the kinetic energies of the colliding gluequarks
($K_V > 2K_\chi$).
The re-annihilation cross section can be written as
\be
\sigma_{{\rm rea},V} = \varepsilon_V \alpha_2 \dfrac{4\pi}{\LDC^{2}}\, ,
\ee 
and the suppression factor $\varepsilon_V$ can be estimated using our semiclassical model by following a procedure similar to the one 
described for the case of glueball emission. 
We find that $\varepsilon_V$ has a behaviour similar to $\varepsilon_\Phi$ as a function of its variables, and a slightly larger absolute value, 
see Fig.~\ref{fig:suppression}.

\subsection{Re-annihilation temperature}

The temperature at which the re-annihilation cross section saturates and the relic abundance freezes out is determined by two competing processes: de-excitation and dissociation. The cross section saturates when the former dominates over the latter for all the bound states with $M_{\Q\Q^*}<2M_\chi$. We will now try to argue that for $T>T_D$ there are states for which the dissociation rate is larger than the de-excitation one. Therefore, the most reasonable scenario is one in which the re-annihilation cross section saturates at $T_R\lesssim T_D$. 

States with $M_{\Q\Q^*}>2M_\chi-M_\Phi$ can be dissociated by glueballs with vanishing kinetic energy. Therefore,  these states are the most easily dissociated since all the glueballs present in the Universe contribute to their breaking rate
\be 
\Gamma_{\rm dis} = n_\Phi \langle \sigma_{\rm dis} v\rangle\,,
\ee
where $\sigma_{\rm dis}\sim \LDC^{-2}$ and $n_\Phi$ is the number density of glueballs which, at $T>T_D$, is dominated by the population coming form the confinement of dark gluons: $n_\Phi\sim T^3$. 
The de-excitation rate can be estimated using the well known result for spontaneous emission 
\footnote{This rate corresponds to dipole transitions and is associated to the usual atomic selection rules. Higher multipole transitions can be considered, but we limit our discussion to the case of the dipole since we are interested only in an order-of-magnitude estimate.}
\be
\Gamma_{\Q\Q^*}\sim \alpha_2 \,\Delta E^3 \left|\langle R_f|\vec{r}|R_i\rangle\right|^2\,,
\ee
where $\Delta E$ is the difference of energy levels. 
A reasonable estimate for this rate can be given for transitions between these states and the ground level. In this case $\Delta E\sim \LDC+\adc^2 M_\Q$, while the matrix element is a fraction of the Bohr radius, $r_b\sim 1/(\adc M_\Q)$. This estimate gives $\Gamma_{\Q\Q^*}$ smaller than $\Gamma_{\rm dis}$ and suggests that for $T>T_D$ re-annihilation cannot proceed through the formation of these states. At $T \sim T_D$ glueballs start to decay. Their number density decays exponentially and the dissociation process becomes soon inefficient. Therefore all the states can contribute to the re-annihilation process and the cross section saturates. 

After the decay of the primordial glueballs, dissociation processes involving electroweak gauge bosons can play a role. However their cross section is suppressed by an electroweak factor and, moreover, their energy distribution is thermal. At $T \lesssim T_{D}$ one needs vector bosons in the tail of the Bose-Einstein distribution in order to have enough energy to dissociate the bound states. As a result, the rate of this process is exponentially suppressed by a factor $\exp[-(2B_\chi -B_{\Q\Q^*})/T]$ and, even if it is efficient at $T_{D}$, it becomes soon inefficient.

For these reasons, we consider the case in which the reannihilation occurs at $T_{D}$ as the most plausible. This is in agreement with what suggested in Ref.~\cite{Boddy:2014qxa}. Due to the large uncertainties on the estimates of the rates, however (especially for what concerns $\Gamma_{\Q\Q^*}$, where neither $\Delta E$ nor the matrix element can be computed from first principles), we do not exclude the possibility that the dissociation processes are never efficient and re-annihilation takes place directly at $\LDC$.

\section{A model with hypercharge}\label{app:LL}

In this article we focused on the minimal block $V$ of Table~\ref{tab:reps} as a benchmark for our analysis. 
However, the $L \oplus \bar{L}$ model has many peculiarities and deserves a separate discussion.
In particular, in this case the DM candidate has non-vanishing hypercharge and interacts at tree level with the $Z$ boson.

\subsection*{Higher-dimensional operators}

This model has a $U(1)_D$ accidental symmetry, comprising dark parity as a subgroup, under which the dark quarks $L$ and $\bar L$ have charge $\pm 1$. 
Differently from the other models, this symmetry is broken by higher-dimensional operators with classical dimension $[\mathcal{O}_{\rm dec}]=5$ of the form
\[
\mathcal{O}_{\rm dec} = \ell \mathcal{G_{\mu\nu}}\sigma^{\mu\nu}L\,.
\]
In order to have a stable DM candidate and make the model viable, one can gauge the $U(1)_D$ in the ultraviolet and break it spontaneously to the dark parity subgroup by means of a scalar field. For instance, if a scalar $\phi$ with charge $2$ acquires a vacuum expectation value the symmetry is broken according to the pattern:
\[
U(1)_D \rightarrow \mathbb{Z}_2.
\]
At the scale of the spontaneous breaking only operators that are dark-parity even are generated, hence the gluequark is absolutely stable.

\subsection*{$Z$-boson mediated direct detection}

Below the confinement scale, the spectrum comprises a composite Dirac fermion with SM quantum numbers $2_{1/2}$, whose EM neutral component is identified with the DM. The non-zero hypercharge induces a tree-level interaction with the $Z$ boson which is strongly constrained by direct searches.
The corresponding spin-independent elastic cross section on nuclei $\mathcal{N}$ is given by~\cite{Goodman:1984dc}:
\[
\sigma = \dfrac{G_{\rm F}^{2} M_{\mathcal{N}}^{2}}{8\pi} \Big(N-(1-4\sin^2\theta_{W})Z \Big)^{2}\, ,
\]
where $Z$ and $N$ are the number of protons and neutrons in the nucleus $\mathcal{N}$ and $M_{\mathcal{N}}$ is its mass.
This cross section is excluded by direct detection experiments for masses $M_{\chi}\lesssim 10^{8}\, \rm GeV$~\cite{1805.12562}. 
This bound rules out the model in the scenario $T_R= T_D$, corresponding to the left panel of Figure~\ref{fig:relic}, but can be satisfied in the 
scenario $T_R = \LDC$. 

In fact, the constraint from direct detection experiments can be also avoided by introducing a heavy singlet gluequark. In this case the presence of Yukawa couplings induces a splitting among the electromagnetically neutral Majorana fermions.  The DM is the lightest among these fermions, so that tree-level elastic scattering mediated by the $Z$ boson cannot exist due to its Majorana nature. Inelastic scatterings are kinematically forbidden if the splitting is large enough; this is easily realized for $M_{\rm N} \lesssim y^2 \times 10^{5} \, \rm TeV$, where $y$ is the Yukawa coupling.  This scenario is analogous to Higgsino DM in supersymmetry, see~\cite{1503.08749,1707.05380} for an extensive discussion.

\subsection*{Accidentally stable mesons}

If the model is not in the conformal window, it is possible to consider the light quark regime. In this case, the model is characterized by the presence of NGBs made of $LL$ or $\bar L\bar L$ constituents which have $U(1)_D$ number $\pm 2$ and therefore cannot decay. If the accidental $U(1)_D$ is gauged in the UV and spontaneously broken to dark parity, then dimension-5 operators can be generated which let the NGBs decay while keeping the gluequarks stable.


\footnotesize

\bibliographystyle{abbrv}

\end{document}